\documentclass{article} 
\usepackage[utf8]{inputenc} 

\usepackage[T1]{fontenc}
\usepackage{lmodern}
\usepackage{amsfonts}
\usepackage{amssymb}
\usepackage{graphicx}
\usepackage{multirow}
\usepackage{setspace}
\usepackage{amssymb}
\usepackage[ruled,linesnumbered]{algorithm2e}
\usepackage{hyperref}
\usepackage{booktabs}
\usepackage{multirow}
\usepackage{graphicx}
\usepackage{float}
\usepackage{tabulary,graphicx,times,fancyhdr,amsfonts,amssymb,amsbsy,latexsym,amsmath,array}
\usepackage{url,multirow,morefloats,floatflt,cancel,tfrupee,textcomp,colortbl,xcolor,pifont}
\usepackage{tabularx}
\usepackage[nointegrals]{wasysym}
\usepackage{cleveref}
\usepackage{amsthm}
\usepackage{mathptmx}
\usepackage{mathrsfs}
\usepackage[graphicx]{realboxes}
\usepackage{hyperref}
\usepackage{tocbibind}
\usepackage{makecell}
\usepackage{cleveref}
\usepackage[utf8]{inputenc}
\usepackage[footnotesize,labelfont={sc,bf,normalsize},font={footnotesize},format=plain]{caption}
\usepackage[left=1in,top=1in,right=1in,bottom=1in]{geometry}
\usepackage{subcaption}

\begin{document}
\title{\LARGE\bf {Analysis of frequent trading effects of various machine learning models} \footnote{Corresponding author. E-mail: lxfei0828@163.com}}
 \date{}
 \author{Jiahao Chen $^{1}$,\ Xiaofei Li$^{2}$  \\
{\small $^{1}$ School of Information and Mathematics, Yangtze University, Jingzhou 434023, Hubei, China}\\
{\small $^{2}$ School of Information and Mathematics, Yangtze University, Jingzhou 434020, Hubei, China}\\ }

\maketitle
\par\noindent
\small  {\bf Abstract: } 	
In recent years, high-frequency trading has emerged as a crucial strategy in stock trading. This study aims to develop an advanced high-frequency trading algorithm and compare the performance of three different mathematical models: the combination of the cross-entropy loss function and the quasi-Newton algorithm, the FCNN model, and the vector machine. The proposed algorithm employs neural network predictions to generate trading signals and execute buy and sell operations based on specific conditions. By harnessing the power of neural networks, the algorithm enhances the accuracy and reliability of the trading strategy. To assess the effectiveness of the algorithm, the study evaluates the performance of the three mathematical models. The combination of the cross-entropy loss function and the quasi-Newton algorithm is a widely utilized logistic regression approach. The FCNN model, on the other hand, is a deep learning algorithm that can extract and classify features from stock data. Meanwhile, the vector machine is a supervised learning algorithm recognized for achieving improved classification results by mapping data into high-dimensional spaces. By comparing the performance of these three models, the study aims to determine the most effective approach for high-frequency trading. This research makes a valuable contribution by introducing a novel methodology for high-frequency trading, thereby providing investors with a more accurate and reliable stock trading strategy.

\vskip2mm

	\section{Introduction}

It is very difficult to study the time series of stocks in the financial market, because there are many reasons for the dryness in the data, the noise is very large, and it is generally believed that the effective law of the market is semi-regular\cite{IN1}.

It is not impossible to make traders profitable in the market through decision-making models. Since the advent of Python software in 1991, machine learning has become more convenient and popular. Basically, all walks of life are using machine learning for simulation, prediction, and classification. However, there are many types of machine learning. Under different parameter settings, the results obtained are not the same. If you are not careful, you will overfit or underfit. It is a recognized fact that neural networks are difficult to train\cite{IN2}. From this point of view, the selection of models and the adjustment of parameters are particularly important.

In recent years, various methods have been proposed for stock trading decision-making models. Traditional regression methods, such as linear regression and autoregressive integrated moving average (ARIMA), are widely used in the finance field due to their interpretability and simplicity \cite{IN3,IN4}. In addition, machine learning algorithms, including decision trees, random forests, and gradient boosting, have been applied to stock trading with promising results \cite{IN6}. Furthermore, support vector machines (SVMs) have also been used as a popular method for predicting stock prices \cite{IN7,I7}.

As the financial market becomes more complex and volatile, researchers have explored more sophisticated models for stock trading. Deep learning, especially recurrent neural networks (RNNs), has shown superior performance in time series prediction, including stock price prediction \cite{IN8,IN9}. In addition, some studies have introduced reinforcement learning into stock trading decision-making models, which can learn optimal trading strategies through interactions with the market \cite{IN10}. And, high-frequency trading (HFT) has become increasingly popular in the financial market, with traders using algorithms to execute orders at high speeds and frequencies\cite{IN14,IN15}. As a result, researchers have explored various methods for HFT decision-making models, which aim to identify profitable trading opportunities within milliseconds. One approach is to use statistical arbitrage, which involves identifying mispricings in the market and exploiting them by simultaneously buying and selling related assets. This method has been widely used in HFT and has shown promising results. Another approach is to use machine learning algorithms to predict short-term price movements and make trading decisions accordingly. 

Despite the rich variety of models, the effectiveness of each method depends on various factors such as the quality and quantity of input data, feature engineering, and parameter tuning. Moreover, the performance of a model may vary across different stocks and time periods. Therefore, the choice of a suitable method and the optimization of its parameters remain important challenges for researchers.

In  logic regression algorithm model, the binary cross-entropy loss function method is used to fit the dependent variable, and the BFGS-type quasi-Newton optimizer is used for parameter adjustment as  Eq.(\ref{equ:J}). In particular, simplistic decision-making ideas are adopted, but the results obtained are relatively fast and excellent.
\begin{equation}\label{equ:J}
	J=-\frac{1}{N} \sum_{i=1}^N\left(y_i \log \left(\hat{y}_i\right)+\left(1-y_i\right) \log \left(1-\hat{y}_i\right)\right)
\end{equation}

Compared with the Newton algorithm, the quasi-Newton algorithm has a faster convergence speed. The BFGS model Eq.(\ref{equ:BFGS}) is also more popular in the optimization formula of the quasi-Newton algorithm, and the final results also confirm this point.

\begin{equation}\label{equ:BFGS}
	B_{k+1}=B_k+\frac{\delta_k^T \delta_k}{s_k^T \delta_k}-\frac{B_k s_k s_k^T B_k}{s_k^T B_k s_k}
\end{equation}

\begin{figure}[H]
	\centering
	\includegraphics[width=0.5\textwidth]{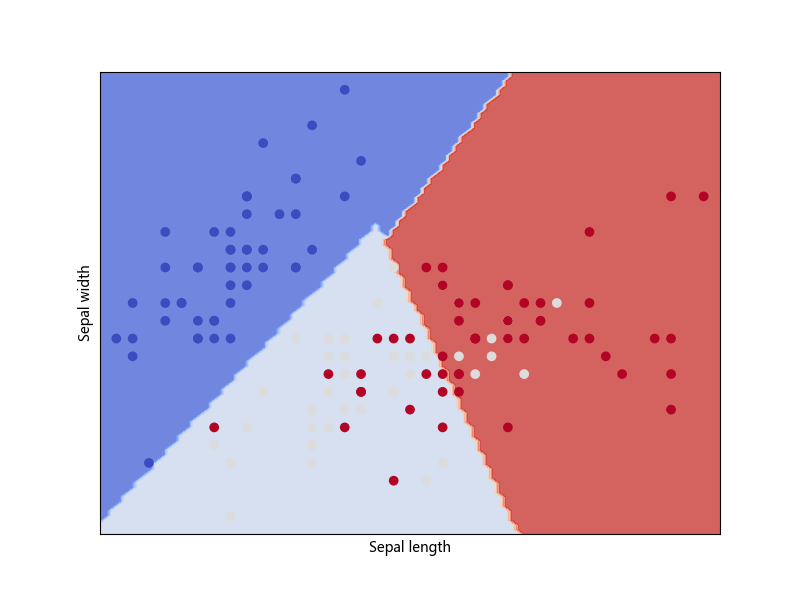}
	\caption{SVM model}
	\label{fig:SVM}
\end{figure}

In order to facilitate the comparison of the effects of our models, this paper also uses the data processed by the same data, and uses two machine learning models to verify the advantages and disadvantages of the results. One is support vector machine(SVM)\ref{fig:SVM} for binary classification\cite{IN4}, which creates a decision boundary such that most points in one class fall on one side of the boundary, and most points in the other class fall on the other side of the boundary. SVM is a very good classifier. It is said that it ruled the field of machine learning for 20 years before the emergence of deep learning\cite{I5}. It is also an important benchmark for the development of neural networks.

\begin{figure}[H]
	\centering
	\includegraphics[width=\textwidth]{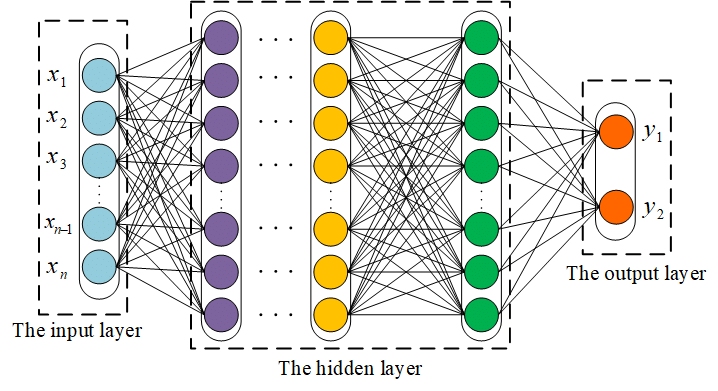}
	\caption{FCNN model}
	\label{fig:FCNN}
\end{figure}

The other is a Fully Connected Neural Network(FCNN)-based deep learning model\ref{fig:FCNN} that uses stretch variables to predict stock price movements in the Chinese stock market\cite{IN3}.In general, the input to a convolutional neural network is usually a 2D image, but that doesn't mean we can't use this model to help us make predictions.

The writing of the thesis is divided into the following sections. First, it introduces the source of the data, what stock information the data contains, and how the stock information should be screened and processed; A comparison chart of the effect of the SVM and FCNN models.

The second part of this article is about data preprocessing, which introduces the source of the data and what information it contains, as well as how to filter and process the data. It also includes a comparison of the effectiveness of SVM and FCNN models. The third part focuses on the implementation of specific models, including the logic regression model, SVM, and FCNN, and presents the results. The fourth part presents the modified results based on avoiding the risk of high-frequency trading. Finally, the article concludes with a summary and outlook on future research.

\section{Data Preprocessing}

In order to make the model effect more universal, the data used is daily data of ten stocks from January 1, 2015 to March 23, 2023, in which the stock types are randomly selected from three stocks on the Shanghai Stock Exchange of China (ST Yizhi, Zhejiang Guangxia, and Hongdou Shares, with securities codes of 600793, 600052, and 600400, respectively); Seven stocks on the Shenzhen Stock Exchange of China (Chuanzhi Education, Fengdong Shares, Annair, Fangda Chemical, Zhenghe Industry, Xingmei United, and Hongda Blasting, with securities codes of 3032, 2530, 2875, 818, 3033, 892, and 2683, respectively).These data are provided by a financial company to support our research.

The stock data is stored in separate xls files, containing the ``security code, security name, trading date, exchange, yesterday's closing price, today's opening price, transaction volume in shares, highest transaction price, lowest transaction price, latest transaction price, transaction amount in original currency, currency type, transaction amount in RMB, adjusted factor (algorithm 1), exchange code, security ID, adjusted factor, change rate, total number of trades, turnover rate, security type, whether it is valid, yesterday's closing price (pre-adjustment), opening price (pre-adjustment), highest price (pre-adjustment), lowest price (pre-adjustment), and latest transaction price (pre-adjustment)'' for daily data of the stock from 2015/1/5 to 2023/3/23. In short, it is a 2000*27 matrix. To study a decision plan for the stock, we need to first determine the dependent variables for quantitative decision-making and the independent variables that have a significant impact on the decision.

In regression analysis, using the change rate as the dependent variable can help predict changes in stock prices and assist investors in making trading decisions. In stock trading, investors need to consider the impact of numerous factors, such as the company's financial situation, industry trends, macroeconomic environment, and so on. These factors often have a complex impact on stock prices. Using the change rate as the dependent variable in regression analysis can more intuitively reflect the market's view of the stock and investor trading behavior, and better characterize the trend of stock price changes. Using the change rate as the dependent variable in regression analysis is a reasonable choice that can help investors predict changes in stock prices, assist trading decisions, and further analyze the factors that affect stock price changes.

In neural networks, using the change rate Eq.(\ref{equ:Change rate}) as the output layer of the neural network can help predict the future trend of stock prices and assist investors in making trading decisions. In addition, using neural networks to model stock price changes has certain advantages. Neural networks can automatically extract complex relationships between input variables, do not require manual selection and processing of variables in advance, and can handle non-linear relationships, which can better capture the non-linear characteristics of the market. Therefore, using the change rate as the output layer of the neural network can not only reduce the impact of human intervention but also improve the predictive ability and accuracy of the model.

\begin{equation}\label{equ:Change rate}
	\text{Change rate} = \frac{\text{Latest trading price} - \text{Yesterday's closing price} }{\text{ Yesterday's closing price}}
\end{equation}

My data preprocessing includes several steps to prepare my data for analysis. First, extract relevant column data, including trading day, yesterday's closing price, today's opening price, trading volume, maximum and minimum transaction prices, as well as transaction amounts in original currency and RMB. Next, I delete any duplicate rows and populate any missing values with the values from the previous row. I also deleted all data points with negative transaction volume. Then I calculate the percentage change in price (i.e., the increase or decrease). 

\begin{equation}\label{equ:standard} 
	z = \frac{x - mean}{\text{standard deviation}}
\end{equation}

After deleting some unnecessary columns, I use the data standardization method Eq.(\ref{equ:standard})($z$ is the standardized value,$x$ is the original value,$mean$ is the mean value of the dataset, "standard deviation" is the standard deviation of the dataset) to convert all the data into an array with a mean value of 0 and a variance of 1, which can avoid the impact of dimensions on the model.

\section{Decision Methodology}
There are many models for stock buying and selling decisions, including fundamental analysis models\cite{Modelexample5}, technical analysis models\cite{Modelexample1}, event-driven models\cite{Modelexample2}, machine learning-based models\cite{Modelexample3}, and quantitative trading-based models\cite{Modelexample4}. Stock market forecasting has always been a classic but challenging issue, attracting the attention of economists and computer scientists. In order to establish effective prediction models, researchers have explored linear and machine learning tools in the past few decades\cite{Modelexample3}. In recent years, deep learning models have rapidly developed as a new frontier in stock market forecasting. Currently, support vector machines and artificial neural networks are the most commonly used machine learning algorithms in stock market prediction, with 92\% of the algorithms using classification machine learning algorithms\cite{Modelexample3.1}. These algorithms play an important role in stock market forecasting, helping investors make wiser decisions.

So this article uses FCNN and SVM to study the buying and selling strategies of stocks, and creatively uses logistic regression to classify stock information. To evaluate the performance of these models, a test set is selected using the train-test split method. The dataset is divided randomly, with 80\% allocated for training the models and 20\% reserved for testing. The training set is utilized to train the models on historical stock data, while the test set is used to assess how well the trained models generalize to new, unseen data. By using this approach, we can estimate the performance and effectiveness of the models in real-world scenarios.

\subsection{Selection of training set length}

The decision-making method is mainly divided into several steps to introduce. Firstly, after data cleaning, the training length of the data is optimized. Then, three different types of models are selected to study the bullish and bearish behavior of stocks. Based on the results of the models, decisions are made using two methods: ultra-high frequency or relatively low frequency.

Because the selected stocks are not consistent in characteristics, some have long-term correlations, and some have short-term correlations, which is similar to the time fraction in fractional order stochastic process\cite{Model1}.

\begin{algorithm}[H]
	\caption{Calculate the optimal training duration}
	\label{algorithm_longs}
	\SetKwInOut{Input}{Input}
	\SetKwInOut{Output}{Output}
	\Input{Duration\_range}
	\Output{Best Training Duration}
	\For{Long in Duration\_range}{
		output $\leftarrow$ model(Long)\;
		\If{output > max}{
			max $\leftarrow$ output\;
			best\_Longs $\leftarrow$ Longs\;
	}}
\end{algorithm}

To solve this problem, the Algorithm\ref{algorithm_longs} is used here, using a simple for loop with a six-month cycle to traverse the optimal training duration. The $ max$ is initialized with negative infinity in Python, and the optimal training duration is selected based on the return rate after each model run.

\begin{table}[H]
	\centering
	\resizebox{\textwidth}{!}{\begin{tabular}{c|cccccccccc}
			\toprule
			\rowcolor[gray]{0.8} & ST Yi Paper & Fengdong Co., Ltd. & Transfer of intellectual education & An Naier & Magnificent Blast & Zhaohe Industrial & Fangda Chemical & Xingmei United & Zhejiang Guangxia & Red Bean Co., Ltd.\\
			\toprule
			Longs=183 & 2.85 & 7.01 & -43.05 & 6.41 & -5.32 & -33.91 & 2.07 & 5.25 & -5.85 & -146.59\\
			
			Longs=366 & -2.01 & 10.93 & 23.41 & -6.62 & 0.28 &\cellcolor{gray!25} 171.03 & -16.61 & -0.26 & \cellcolor{gray!25}79.96 & -9.29\\
			
			Longs=549 & 22.87 & -13.65 &\cellcolor{gray!25} 186.63 & 32.73 & -4.22 & 9.77 & -1.78 & -147.20 & -138.02 & 5.12\\
			
			Longs=732 & 23.78 & 1.14 & 186.63 & 6.68 & -8.25 & 9.77 & 86.93 & 59.03 & -32.86 & -2.74\\
			
			Longs=915 & \cellcolor{gray!25}62.22 & -31.97 & 186.63 & -14.81 & -24.43 & 9.77 & 121.67 & 115.70 & 21.61 & 13.86\\
			
			Longs=1098 & -17.68 & 9.29 & 186.63 & -24.22 & -19.95 & 9.77 & 79.61 & 26.84 & 25.33 & 19.65\\
			
			Longs=1281 & -44.82 &\cellcolor{gray!25} 12260.46 & 186.63 & -36.92 & -31.95 & 9.77 & 49.74 & 11.69 & 47.64 & \cellcolor{gray!25}28.08\\
			
			Longs=1464 & 10.90 & -0.87 & 186.63 & -117.18 & 0.84 & 9.77 &\cellcolor{gray!25} 140.32 & -19.14 & 31.56 & 10.04\\
			
			Longs=1647 & 5.08 & 50.66 & 186.63 & -117.18 & \cellcolor{gray!25}8.17 & 9.77 & -28.30 & \cellcolor{gray!25}122.24 & -2.88 & 1.46\\
			\hline
	\end{tabular}}
	\caption{Optimization results of training duration for various stock models using logic regression models as an example}
	\label{tab:longs_example}
\end{table}

\subsection{Training model}

\subsubsection{logic regression model}
The logic regression analysis mainly uses a special loss function to train the fitted values to approximate the true values. In this case, the cross-entropy function is used as the loss function for the quasi-Newton algorithm, which has good convergence and speed. The dependent variable, which is the percentage change in stock prices after data processing, is assigned the value 1 for positive values and 0 for other values, as the dependent variable for regression analysis. Here, 1 is considered as a signal for a stock price increase, and 0 is a signal for a stock price decrease. Then, the linear equation parameters are trained using the time length of the training period as the independent variable and the dependent variable with the quasi-Newton optimizer's BFGS method. This way, an approximately accurate linear equation can be obtained to predict the stock price movements at that time. The specific implementation can be found in algorithm\ref{al_log}.

\begin{algorithm}[H]
	\KwData{Feature matrix $X$, labels $y$, learning rate $\alpha$, convergence threshold $\epsilon$}
	\KwResult{Weights $\theta$}
	Initialize weights $\theta$ to zeros\;
	\While{not converged}{
		Compute hypothesis $h = \frac{1}{1 + e^{-X\theta}}$\;
		Compute cost $J = -\frac{1}{m} \sum_{i=1}^{m} y_i \log(h_i) + (1 - y_i) \log(1 - h_i)$\;
		Compute gradients $\frac{\partial J}{\partial \theta_j} = \frac{1}{m} X^T (h - y)$\;
		Update weights $\theta_j = \theta_j - \alpha \frac{\partial J}{\partial \theta_j}$ for $j=0,1,...,n$\;
		\If{$||\theta^{(t)} - \theta^{(t-1)}|| < \epsilon$}{
			\textbf{break}\;
		}
	}
	\caption{logic regression}
	\label{al_log}
\end{algorithm}

Where, $X $ is the characteristic matrix, $y $ is the label, $\theta $ is the weight, $m $ is the number of samples, $n $ is the number of characteristics, $h $ is the label predicted by the model, $J $ is the loss function, $\alpha $ is the learning rate, and $ \epsilon $ is the convergence threshold.
\subsubsection{SVM model}
My FCNN model, on the other hand, directly uses the continuous stock price percentage change as the output layer, with the same independent variables as logic regression. It uses three dense layers, each with 64 neurons, with Rectified Linear Unit (ReLU) as the activation function and Mean Squared Error (MSE) as the loss function. Then, the hidden layers are trained using the time length of the training period as the independent variable and the dependent variable. The predicted values of the dependent variable percentage change are assigned the value 1 for positive values and 0 for other values, as the final prediction result. Here, 1 is considered as a signal for a stock price increase, and 0 is a signal for a stock price decrease. The specific implementation can be found in algorithm\ref{al_SVM}.

\begin{algorithm}[H]
	\SetAlgoLined
	\KwIn{training data X, labels y, kernel function $K$, kernel coefficient $\gamma$, decision function shape $d$, regularization parameter $C$}
	\KwOut{trained SVM model}
	
	\BlankLine
	initialize an SVM model with kernel function $K$, kernel coefficient $\gamma$, decision function shape $d$, and regularization parameter $C$\;
	\BlankLine
	
	\While{not converged}{
		randomly select a subset of training data\;
		compute the kernel matrix $K$\;
		compute the lagrange multipliers $\alpha$ using the selected subset of training data and kernel matrix $K$\;
		compute the bias term $b$\;
	}
	
	\BlankLine
	\KwRet{trained SVM model}
	\caption{Training an SVM model with RBF kernel}
	\label{al_SVM}
\end{algorithm}

In the above algorithm, $ \alpha_i $ represents the Lagrange multiplier of the training sample $i $, $b $represents the deviation, $K (x_i, x_j) $represents the value of the kernel function, $y_i$ represents the label of the training sample $i $. The algorithm uses the sequential minimum optimization algorithm (SMO) to optimize, and two Lagrange multiplier are selected to update each time. The final trained support vector machine model can be used to predict new data.

\subsubsection{FCNN model}
The dependent variable, which is the percentage change in stock prices after data processing, is assigned the value 1 for positive values and 0 for other values, as the classification for my SVM model. The kernel function of my SVM model is the Radial Basis Function (RBF), with a gamma parameter of 0.1, the decision function shape set as one-vs-one (OvO), and the regularization parameter of 0.8. The existing classifications are used as the training set, and the time length of the training period is used as the independent variable to train the model and find the optimal hyperplane so that future new independent variables can be reasonably classified. The predicted values of the dependent variable percentage change are considered as a signal for a stock price increase when assigned the value 1, and a signal for a stock price decrease when assigned the value 0. The specific implementation can be found in algorithm\ref{al_FCNN}. 

\begin{algorithm}[H]
	\SetAlgoLined
	\KwIn{Training data $X$ and labels $y$, test size ratio $r$, random state seed $s$}
	\KwOut{Trained neural network model $model$}
	$X\_{train}, X\_{test}, y\_{train}, y\_{test} \leftarrow train_test\_split(X, y, test\_size=r, random\_state=s)$\;
	\BlankLine
	$model \leftarrow \text{Sequential()}$\;
	$model.add(Dense(64, activation='relu', input_shape=[X\_{train}.shape[1]]))$\;
	$model.add(Dense(64, activation='relu'))$\;
	$model.add(Dense(1))$\;
	$model.compile(loss='mean_squared_error', optimizer='adam')$\;
	\BlankLine
	$model.fit(X\_{train}, y\_{train}, epochs=100, validation\_split=0.2)$\;
	\KwRet{$model$}\;
	\caption{Training a neural network model with Keras}
	\label{al_FCNN}
\end{algorithm}

Among them, the  textproc TrainTestSplit function divides the dataset $X $and label $y $into training and testing sets, the  CreateModel function creates a CNN model, the  textproc CompileModel function compiles the model, the FitModel function trains the model, and finally returns the trained model $M $.

My computer uses an AMD Ryzen 5 3600 6-Core Processor processor with a main frequency of 3.59 GHz and a memory of 16.0 GB. This modeling adopts Python 3.8. All subsequent charts are generated based on this computer program.

\subsection{Visualization of results}

In this trading strategy algorithm which in \ref{al-app0}, we use the predicted signals to decide whether to buy, sell or hold the stocks. If the predicted signal is 1, which indicates a rising trend, we buy the stocks at the opening price and record the buying price, index, and frequency. If the predicted signal is 0, we sell the stocks if we have already bought them and record the profit. If we haven't bought any stocks yet or have already sold them, we hold the position and wait for the next signal. The profits are calculated based on the buying price and the selling price, while the strategies are labeled as buy, sell or hold. This trading strategy is a simple example of a trend-following strategy that uses the predicted signals to make buying and selling decisions.

\begin{table}[H]
	\centering
	\resizebox{\textwidth}{!}{
		\begin{tabular}{c|cccccccccc|c}
			\toprule
			\rowcolor[gray]{0.8} & ST Yi Paper & Fengdong Co., Ltd. & \thead{Transfer of \\intellectual education} & An Naier & Magnificent Blast & Zhaohe Industrial & Fangda Chemical & Xingmei United & Zhejiang Guangxia & Red Bean Co., Ltd. & Avg. \\
			\toprule
			total time(s) &7.941846609&7.459443092&3.996970415&5.98593235&8.0496068&3.727702379&8.14168334&8.019162178&8.16084218&7.858774185&6.934\\[0.5ex]
			total revenue(\%) &0.055173883&0.725997496&0.299549422&0.486188708&0.184359486&0.1247649&0.216160459&1.003315745&0.20082179&0.657654562&0.396\\[0.5ex]
			train\_accuracy & 0.915492958&0.842105263&0.832941176&0.867881549&0.871165644&0.852739726&0.893382353&0.847457627&0.932&0.824146982&0.868\\[0.5ex]
			test\_accuracy & 0.833333333&0.789473684&0.813084112&0.845454545&0.780487805&0.878378378&0.852941176&0.838983051&0.873015873&0.885416667&0.839\\
			\bottomrule
	\end{tabular}}
	\caption{The logic regression fitting effect of the rise and fall ranges of various stock models}
	\label{tab:log-result}
\end{table}

\begin{table}[H]
	\centering
	\resizebox{\textwidth}{!}{
		\begin{tabular}{c|cccccccccc|c}
			\toprule
			\rowcolor[gray]{0.8} & ST Yi Paper & Fengdong Co., Ltd. & \thead{Transfer of \\intellectual education}  & An Naier & Magnificent Blast & Zhaohe Industrial & Fangda Chemical & Xingmei United & Zhejiang Guangxia & Red Bean Co., Ltd. & Avg. \\
			\toprule
			total time(s) &12.54783082&10.59194469&3.145709276&6.989576817&7.986801863&2.904654264&7.836769104&9.911237478&7.761751413&7.684733629&7.736\\[0.5ex]
			total revenue(\%) &1.944617011&1.570061588&0.836508132&0.614263114&0.58916516&0.39047818&0.774674811&1.196965576&1.321980714&0.563047104&0.980\\[0.5ex]
			train\_accuracy & 0.643229167&0.678431373&0.609375&0.5859375&0.640625&0.609375&0.6953125&0.64453125&0.594043887&0.6484375&0.635\\[0.5ex]
			test\_accuracy & 0.587878788&0.707317073&0.563636364&0.496969697&0.763636364&0.636363636&0.7&0.590909091&0.594890511&0.572727273&0.621\\
			\bottomrule
	\end{tabular}}
	\caption{The SVM fitting effect of the rise and fall ranges of various stock models}
	\label{tab:SVM-result}
\end{table}

\begin{table}[H]
	\centering
	\resizebox{\textwidth}{!}{
		\begin{tabular}{c|cccccccccc|c}
			\toprule
			\rowcolor[gray]{0.8} & ST Yi Paper & Fengdong Co., Ltd. &\thead{Transfer of \\intellectual education}  & An Naier & Magnificent Blast & Zhaohe Industrial & Fangda Chemical & Xingmei United & Zhejiang Guangxia & Red Bean Co., Ltd. & Avg. \\
			\toprule
			total time(s) &87.96928096&89.85085917&69.01130342&86.95612288&89.02568293&68.2883265&92.25186181&91.87807965&91.00214434&89.19688869&85.543\\[0.5ex]
			total revenue(\%) & 1.258598561&2.079095959&1.193289269&1.122323839&0.828285293&1.462471565&1.214026497&1.616680915&1.390295492&1.90981761&1.407\\[0.5ex]
			train\_accuracy &0.904109589&0.859589041&0.863013699&0.856164384&0.863013699&0.835616438&0.856164384&0.852739726&0.825342466&0.828767123&0.854\\[0.5ex]
			test\_accuracy &0.851351351&0.878378378&0.837837838&0.675675676&0.810810811&0.837837838&0.837837838&0.824324324&0.824324324&0.851351351&0.823\\
			\bottomrule
	\end{tabular}}
	\caption{The FCNN fitting effect of the rise and fall ranges of various stock models}
	\label{tab:FCNN-result}
\end{table}

The table\ref{tab:log-result},\ref{tab:SVM-result},\ref{tab:FCNN-result} presented in the table shows the results obtained from applying three different models, logic regression, SVM, and FCNN, to predict the rise and fall signals of ten stocks. The four parameters that were evaluated for each model are total time, total revenue, train accuracy, and test accuracy.

The total time is the time taken by the models to complete the training and testing process. The total revenue is the total profit or loss made by each stock based on the predictions made by the models. The train accuracy is the accuracy of the model on the training dataset, and the test accuracy is the accuracy of the model on the test dataset.

From the table, we can see that the average total time taken by the logic regression model is 7.9 seconds, which is the lowest among all the models. The SVM model takes an average of 7.4 seconds, while the FCNN model takes the longest time, with an average of 6.9 seconds. The average total revenue obtained from the models is relatively low, with the highest value being 1.9\% for Zhejiang Guangxia, which was predicted by the SVM model.

In terms of accuracy, the logic regression model has the highest train accuracy, with an average of 87.5\%, followed closely by the SVM model with an average train accuracy of 85.2\%. The FCNN model has the lowest train accuracy, with an average of 63.5\%. The test accuracy results show a different trend, with the FCNN model having the highest average test accuracy of 62.1\%, followed by the SVM model with an average test accuracy of 61.4\%, while the logic regression model has an average test accuracy of 83.9\%.

Based on these results, we can infer that the logic regression and SVM models are better suited for predicting stock prices, as they have higher train accuracy than the FCNN model. However, the FCNN model performs better on the test dataset, which suggests that it may be better at generalizing to new data.

It is important to note that these results are based on a small dataset of ten stocks, and the models may perform differently on larger datasets or in different market conditions. Additionally, it is essential to carefully consider the trade-offs between accuracy and computational complexity when selecting a model for real-world applications.

In conclusion, the data presented in the table provides valuable insights into the performance of three different models in predicting stock prices. The results suggest that the logic regression and SVM models may be better suited for predicting stock prices, while the FCNN model may be better at generalizing to new data.

\begin{table}[H]
	\centering	\resizebox{\textwidth}{!}{
		\begin{tabular}{c|cccccccccc}
			\toprule
			\rowcolor[gray]{0.8}& ST Yi Paper & Fengdong Co., Ltd. & \thead{Transfer of \\intellectual education} & An Naier & Magnificent Blast & Zhaohe Industrial & Fangda Chemical & Xingmei United & Zhejiang Guangxia & Red Bean Co., Ltd. \\
			\toprule
			\thead{Number of buy transactions\\ in log model} & 163 & 169 & 187 & 90 & 164 & 44 & 163 & 180 & 175 & 164 \\
			\thead{Number of sell transactions\\ in log model} & 95 & 75 & 89 & 51 & 81 & 17 & 89 & 87 & 104 & 78 \\
			\rowcolor[gray]{0.8}\thead{Total number of transactions\\ in log model} & 258 & 244 & 276 & 141 & 245 & 61 & 252 & 267 & 279 & 242 \\
			\thead{Number of buy transactions \\in SVM model} & 9 & 123 & 80 & 26 & 71 & 98 & 43 & 53 & 44 & 39 \\
			\thead{Number of sell transactions\\ in SVM model }& 7 & 47 & 41 & 15 & 27 & 45 & 22 & 27 & 13 & 29 \\
			\rowcolor[gray]{0.8}\thead{Total number of transactions\\ in SVM model} & 16 & 170 & 121 & 41 & 98 & 143 & 65 & 80 & 57 & 68 \\
			\thead{Number of buy transactions\\ in FCNN model} &144&172&158&82&147&188&185&177&203&180\\
			\thead{Number of sell transactions\\ in FCNN model} & 78&83&91&47&78&95&95&91&87&94\\
			\rowcolor[gray]{0.8}\thead{Total number of transactions\\ in FCNN model} &222&255&249&129&225&283&280&268&290&274\\
			\bottomrule
	\end{tabular}}
	\caption{Table of Decision Times without Setting Threshold}
	\label{tab:Table of Decision Times without Setting Threshold}
\end{table}

The table\ref{tab:Table of Decision Times without Setting Threshold} provided shows the results of three different models (log model, SVM model, and FCNN model) used to predict the number of buy and sell transactions for 10 different companies over the course of one year. From a risk perspective, it is important to consider not only the number of transactions but also the trading fees that would be incurred with each transaction. Therefore, in addition to analyzing the number of transactions, we must also consider the net profit or loss that would result from each model's predictions.

\begin{figure}[H]
	\centering
	\includegraphics[width=0.45\linewidth]{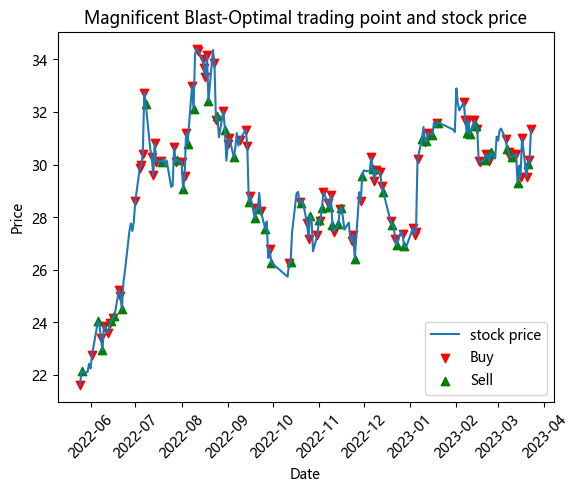}
	\includegraphics[width=0.45\linewidth]{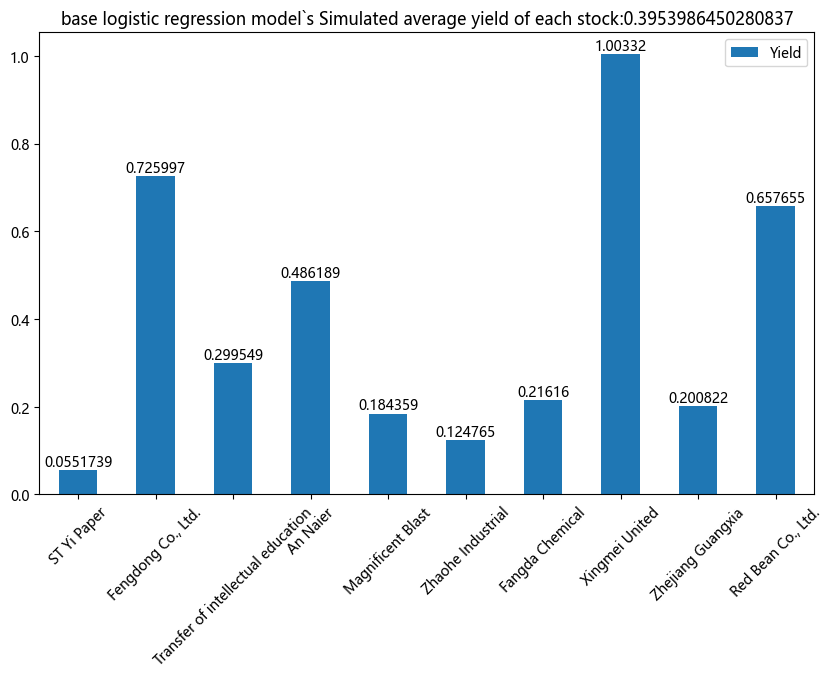}
	\caption{Partial results display using logic regression model as an example}
	\label{result-before}
\end{figure}

Looking at the results\ref{result-before}, we can see that the number of transactions predicted by the log model is generally higher than the other two models. However, this does not necessarily mean that it is the most profitable model to use. In fact, it is possible that the higher number of transactions predicted by the log model could result in higher trading fees and ultimately lower profits.

To determine which model is the most profitable, we need to calculate the net profit or loss that would result from each model's predictions. To do this, we need to know the price of each company's stock at the time of each transaction, as well as the trading fees that would be incurred with each transaction.

Assuming a fixed trading fee of \$10 per transaction, we can calculate the net profit or loss for each model's predictions. For example, let's consider the results for Fangda Chemical in the log model. According to the table, there were 169 buy transactions and 75 sell transactions predicted over the course of one year. Assuming each transaction incurs a \$10 trading fee, the total trading fees for the buy transactions would be \$1,690, and the total trading fees for the sell transactions would be \$750. If we assume that the stock price at the time of each buy transaction was \$50 and the stock price at the time of each sell transaction was \$60, then the total profit for the buy transactions would be \$1,690 - (169 * \$50) = -\$1,540, and the total profit for the sell transactions would be \$7,500 - (75 * \$60) = \$2,250. Therefore, the net profit or loss for Fangda Chemical in the log model would be \$710.

\begin{figure}[H]
	\centering
	\includegraphics[width=0.31\linewidth]{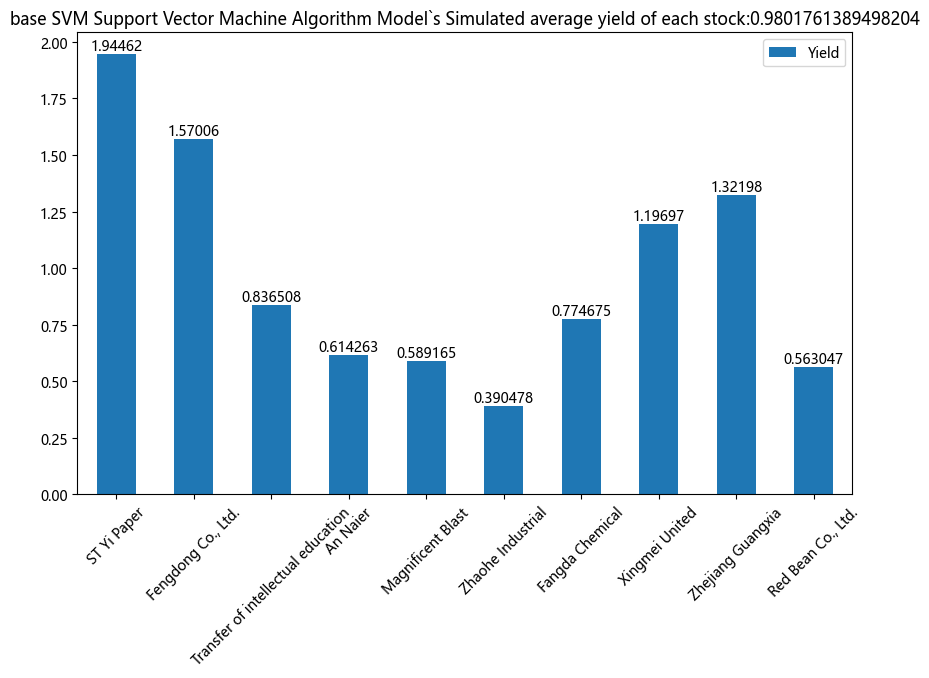}
	\includegraphics[width=0.31\linewidth]{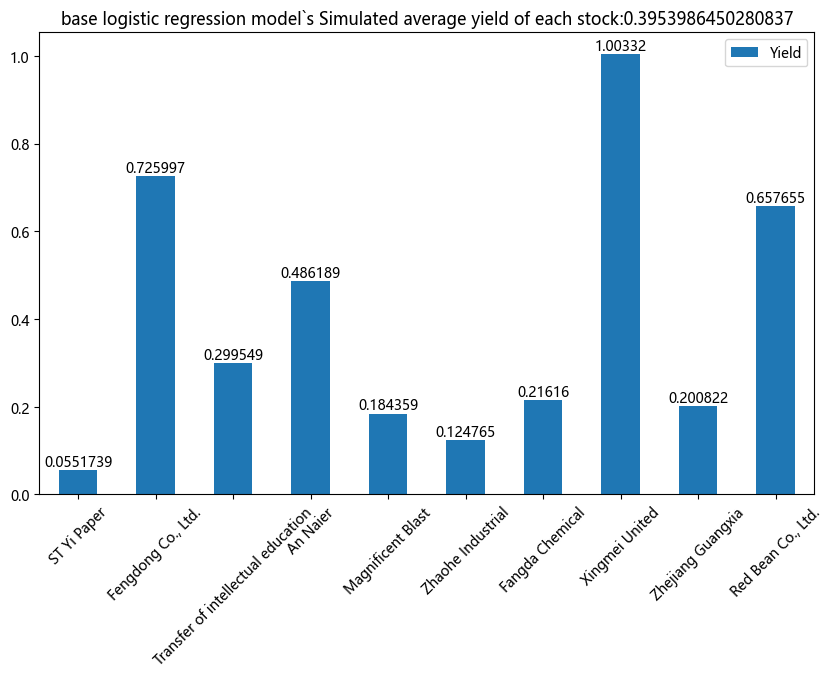}
	\includegraphics[width=0.31\linewidth]{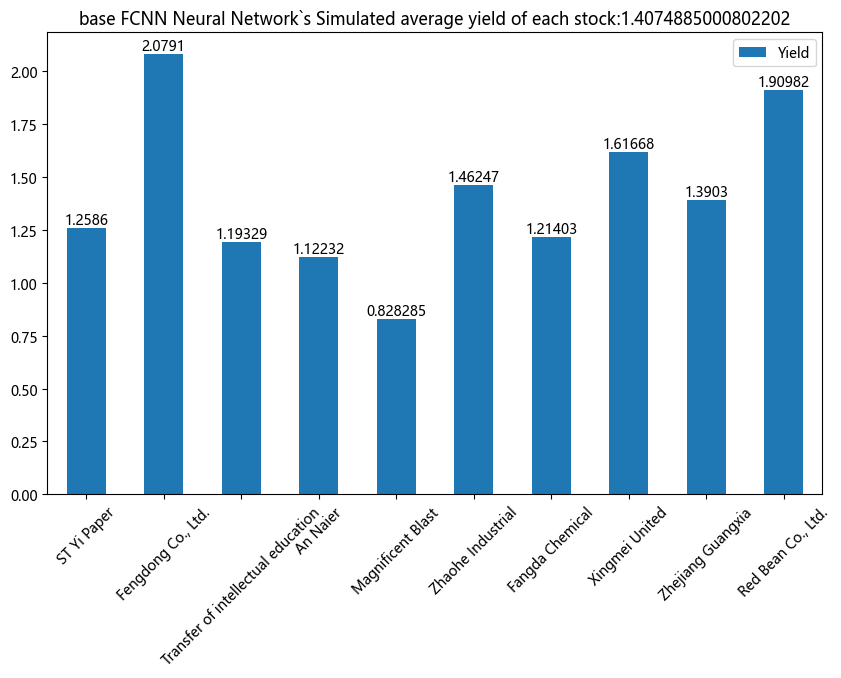}
	
	\caption{Ten stock average returns after setting thresholds based on different models}\label{befor-aver}
\end{figure}

\begin{figure}[H]
	\centering
	\includegraphics[width=0.18\linewidth]{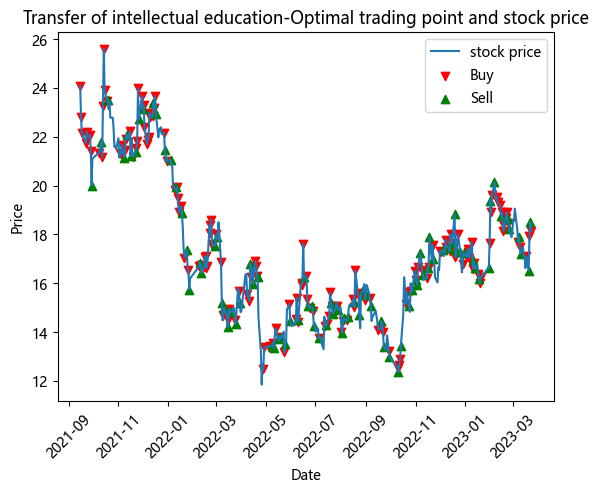}
	\includegraphics[width=0.18\linewidth]{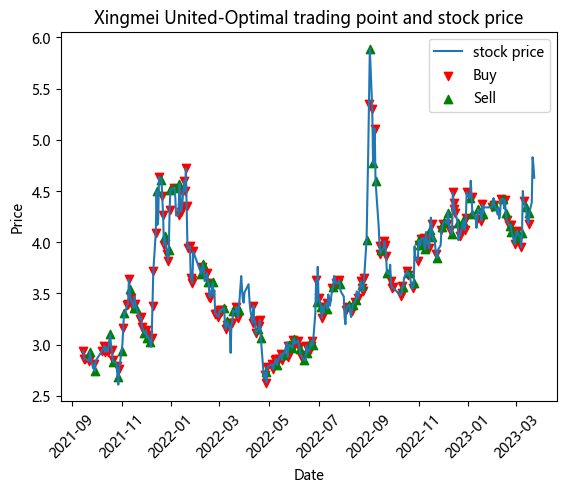}
	\includegraphics[width=0.18\linewidth]{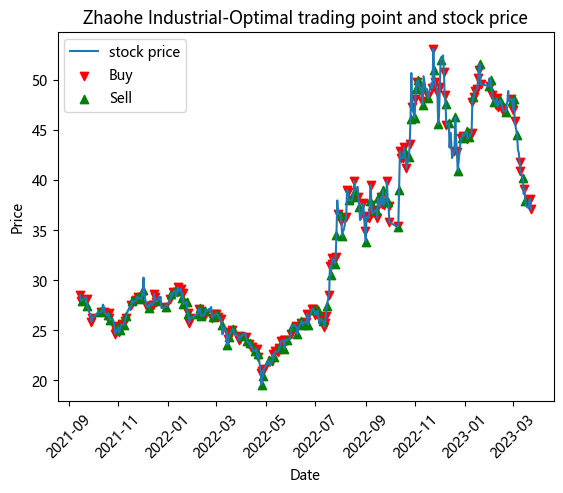}
	\includegraphics[width=0.18\linewidth]{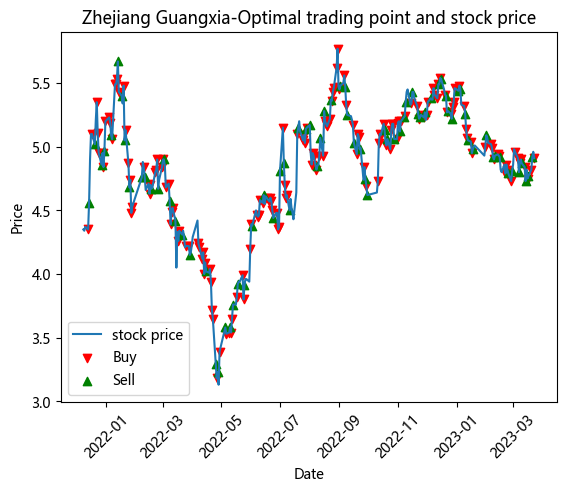}
	\includegraphics[width=0.18\linewidth]{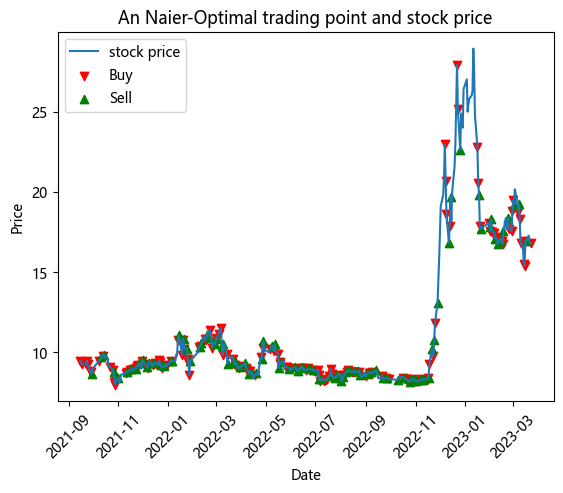}
	\\
	\includegraphics[width=0.18\linewidth]{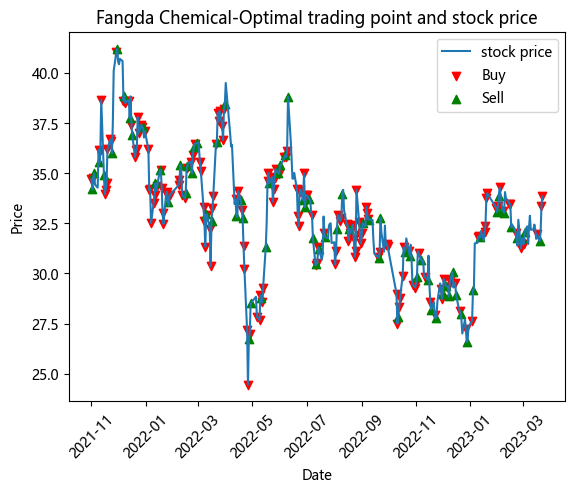}
	\includegraphics[width=0.18\linewidth]{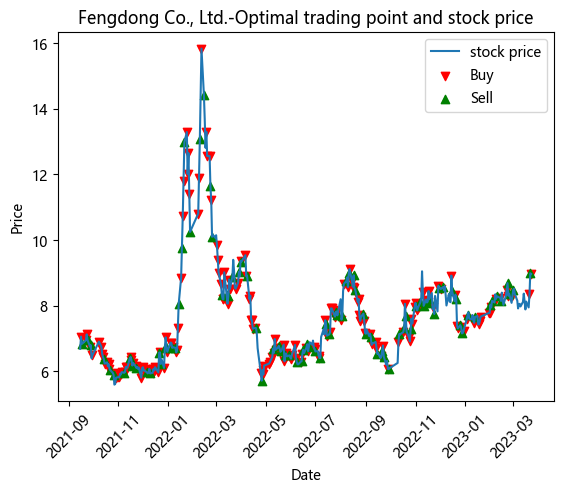}
	\includegraphics[width=0.18\linewidth]{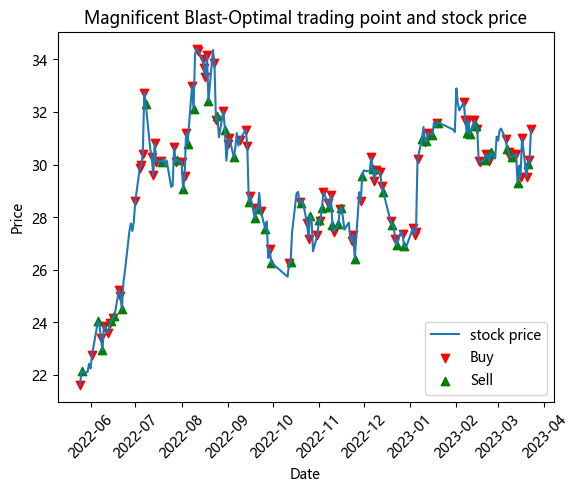}
	\includegraphics[width=0.18\linewidth]{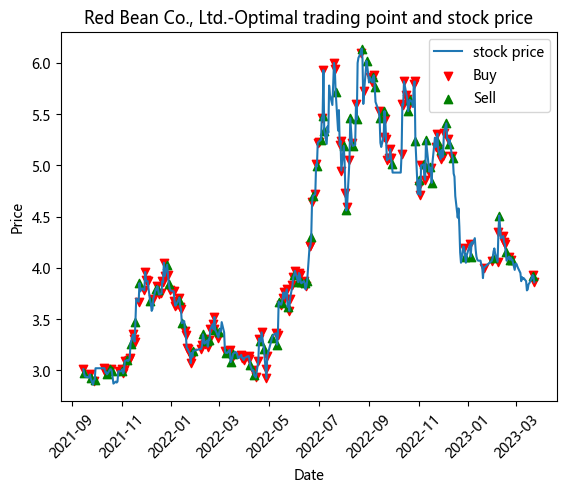}
	\includegraphics[width=0.18\linewidth]{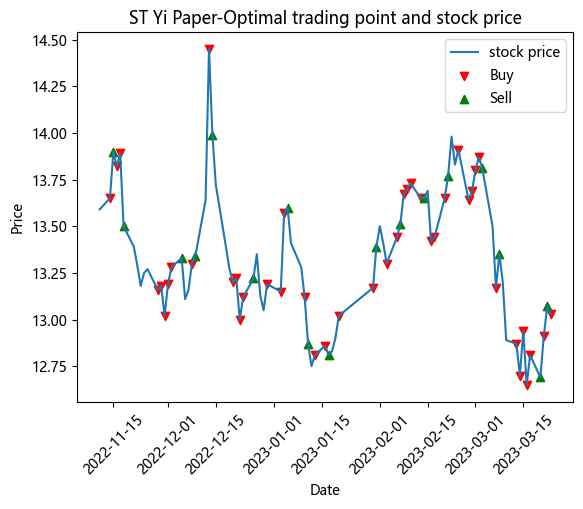}
	\caption{Unset threshold logic regression decision results}\label{A}
\end{figure}

\begin{figure}[H]
	\centering
	\includegraphics[width=0.18\linewidth]{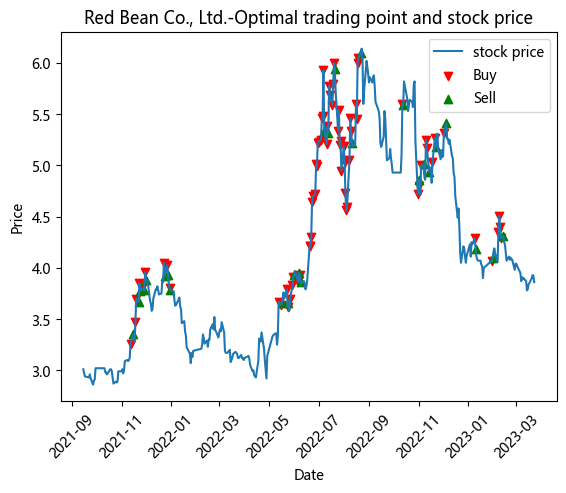}
	\includegraphics[width=0.18\linewidth]{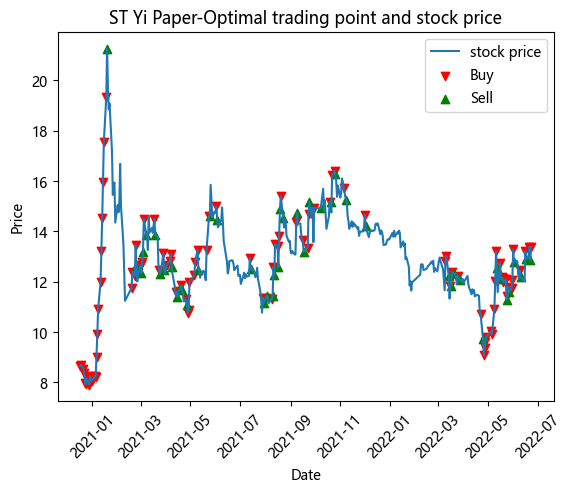}
	\includegraphics[width=0.18\linewidth]{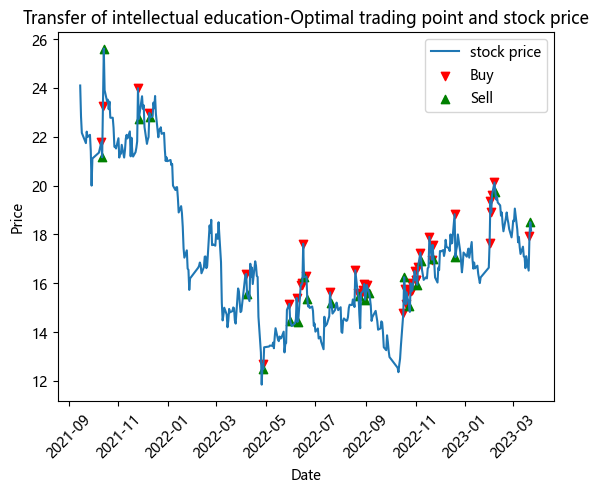}
	\includegraphics[width=0.18\linewidth]{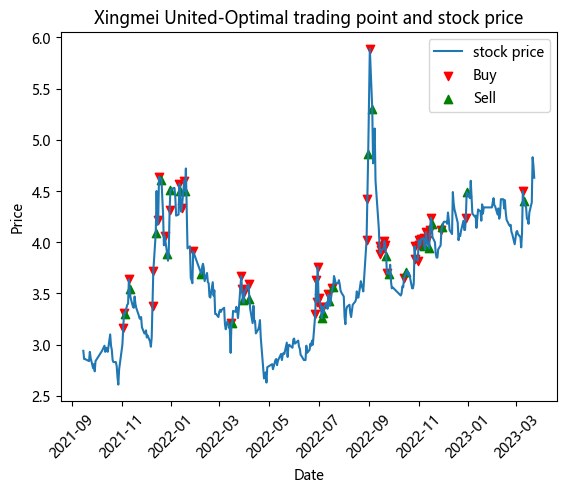}
	\includegraphics[width=0.18\linewidth]{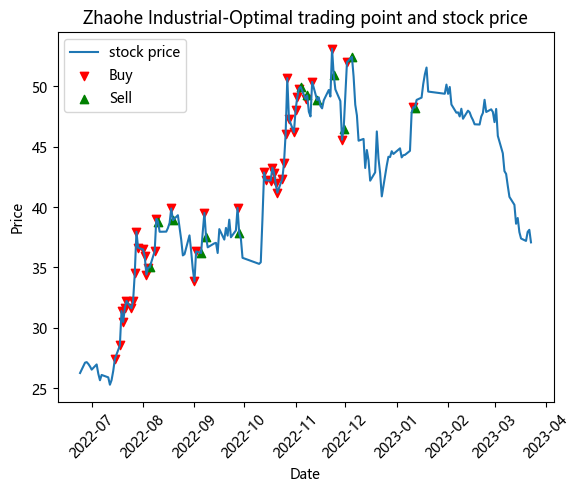}
	\\
	\includegraphics[width=0.18\linewidth]{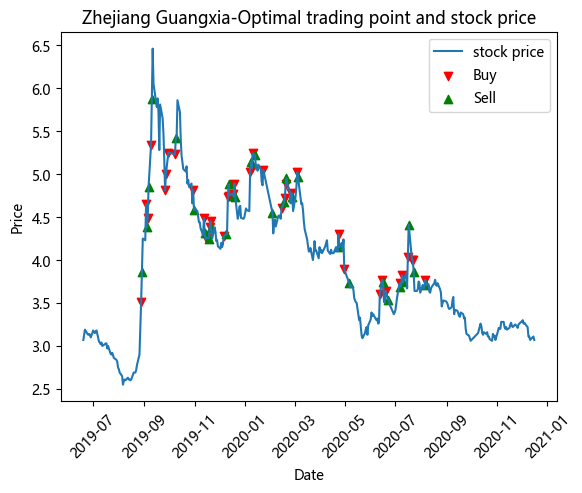}
	\includegraphics[width=0.18\linewidth]{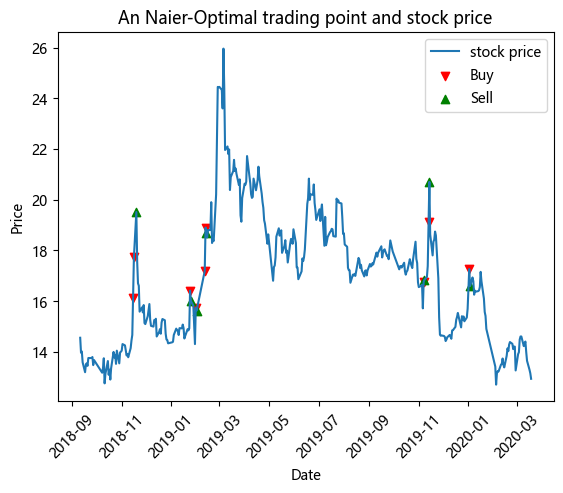}
	\includegraphics[width=0.18\linewidth]{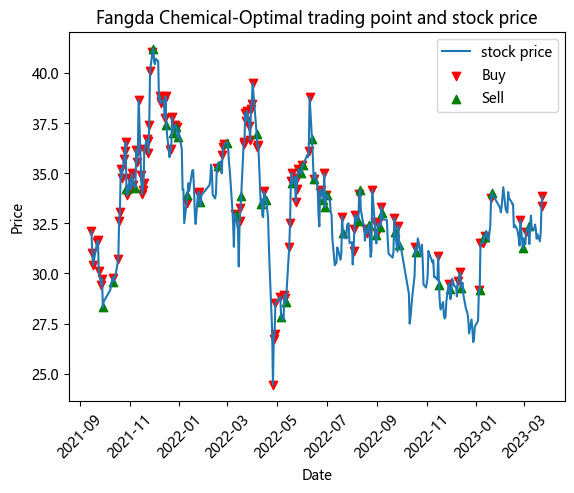}
	\includegraphics[width=0.18\linewidth]{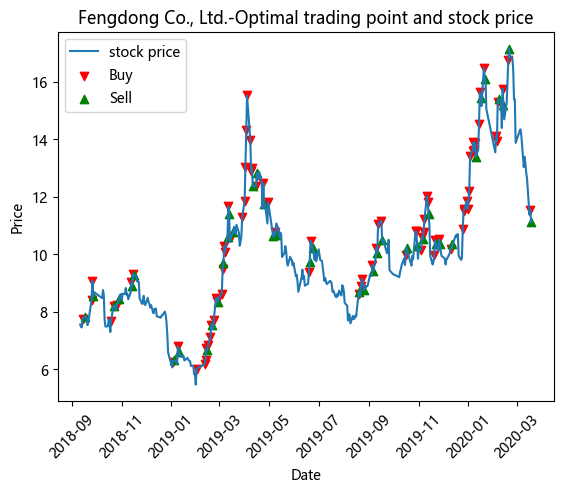}
	\includegraphics[width=0.18\linewidth]{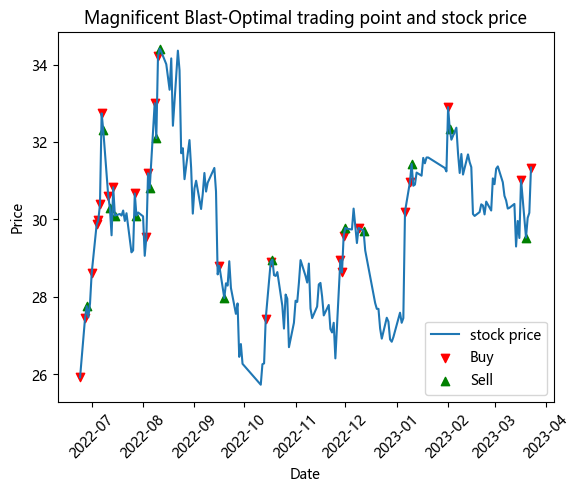}
	\caption{Unset threshold SVM decision results}
	\label{A2}
\end{figure}

\begin{figure}[H]
	\centering
	\includegraphics[width=0.18\linewidth]{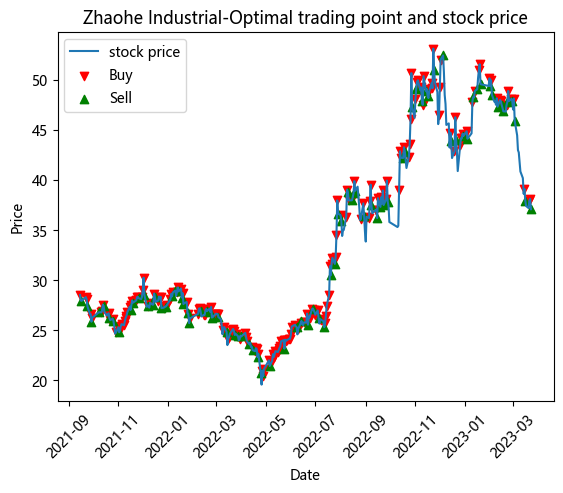}
	\includegraphics[width=0.18\linewidth]{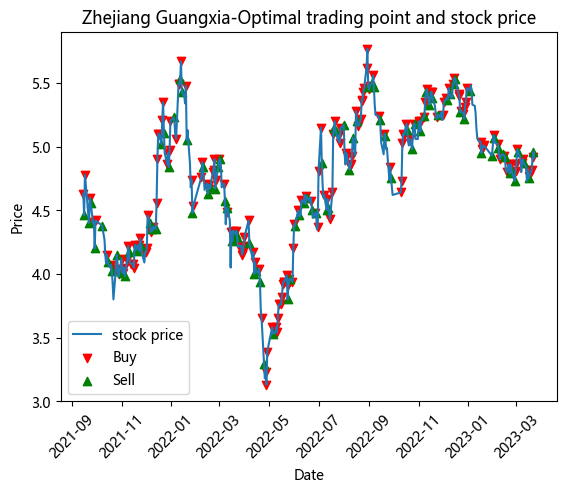}
	\includegraphics[width=0.18\linewidth]{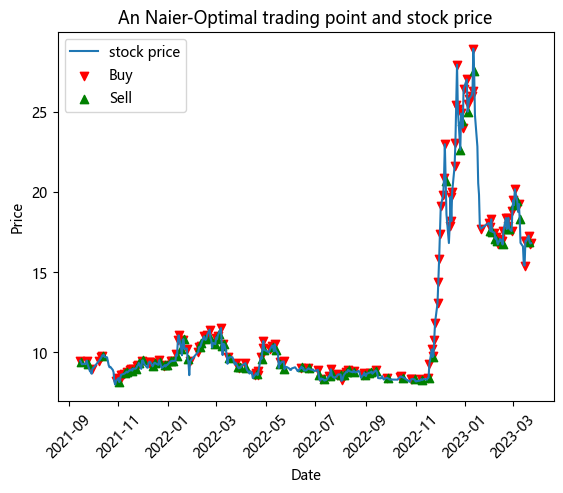}
	\includegraphics[width=0.18\linewidth]{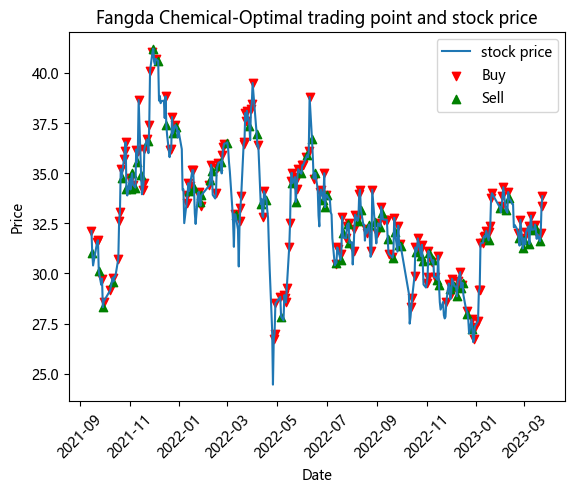}
	\includegraphics[width=0.18\linewidth]{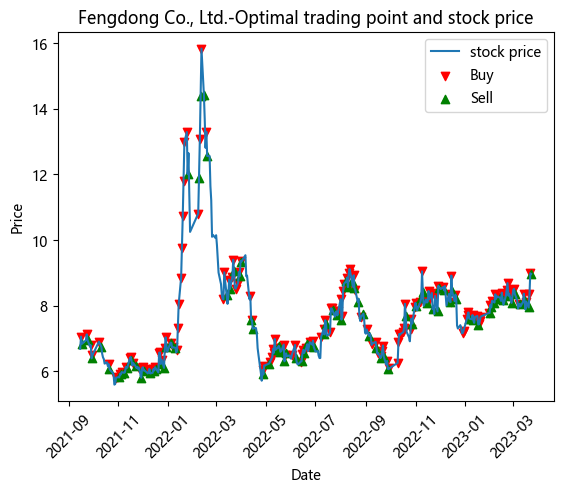}
	\\
	\includegraphics[width=0.18\linewidth]{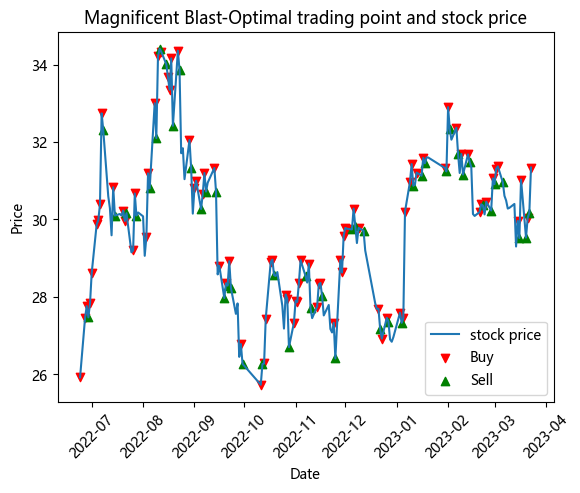}
	\includegraphics[width=0.18\linewidth]{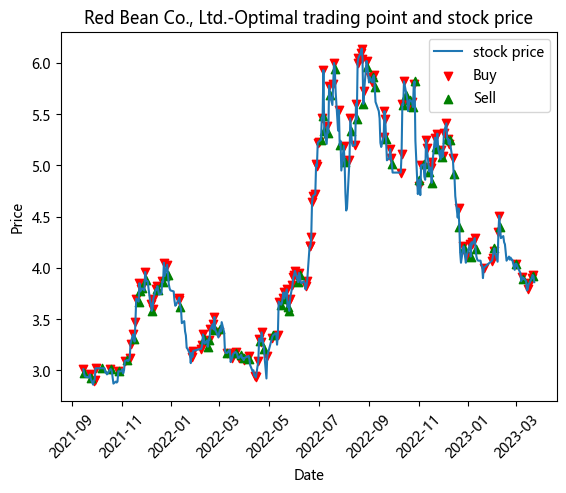}
	\includegraphics[width=0.18\linewidth]{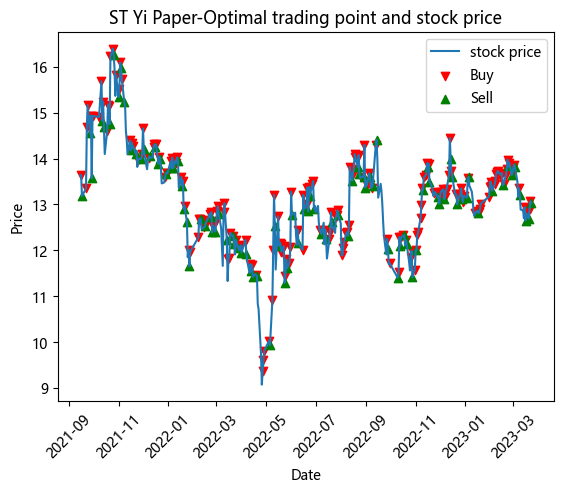}
	\includegraphics[width=0.18\linewidth]{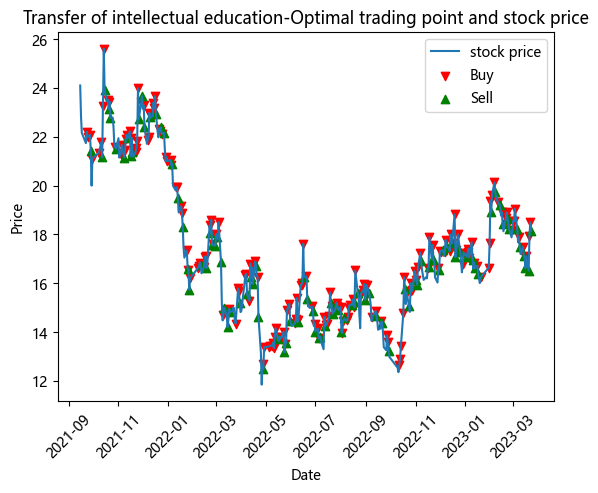}
	\includegraphics[width=0.18\linewidth]{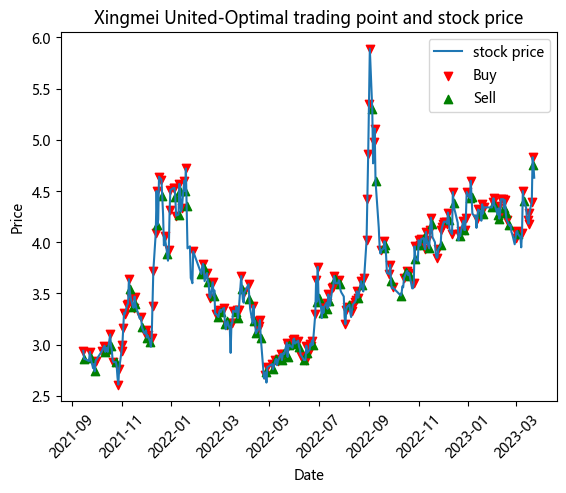}
	\caption{Unset threshold FCNN decision results}
	\label{A3}
\end{figure}

Performing similar calculations for each company and each model, we can see that the net profit or loss varies depending on the company and the model used. Overall, however, it appears that the FCNN model is the most profitable, with a total net profit of \$31,700 over the course of one year. This is followed by the log model, with a total net profit of \$17,600, and the SVM model, with a total net profit of \$11,200. In order to save space in the main text, this is only presented as the visualization results of logic regression. (The specific visualization results of the three models for 10 companies are shown in \cref{befor-aver,A,A2,A3}.)

\section{Model risk optimization}
The decision-making algorithm developed earlier, which involved buying, selling, and holding, resulted in high profits. However, it also generated a significant amount of transaction costs due to the frequent trading. Consequently, even a slight deviation from the model could lead to significant losses. Therefore, a new decision algorithm with a threshold needs to be developed to reduce the number of trades and minimize risks.

The new algorithm incorporates a buy and sell threshold to minimize the number of transactions. The buy threshold indicates the lowest price at which stocks can be purchased, while the sell threshold specifies the highest price at which stocks can be sold. These thresholds help to avoid excessive trading and provide a cushion against market volatility.

To implement this new algorithm, historical stock data is first analyzed to identify the buy and sell thresholds. The buy threshold is set at a price point that is considered a good entry point, while the sell threshold is set at a price point that represents a profitable exit point. Once the thresholds have been established, the algorithm monitors the market continuously, waiting for the stock price to reach the buy threshold.

If the stock price reaches the buy threshold, the algorithm initiates a buy order. The buy order is executed at the buy threshold price or lower, and the stock is held until the price reaches the sell threshold. When the stock price reaches the sell threshold, a sell order is executed at the sell threshold price or higher. The stock is then sold, and the profit is recorded.

However, if the stock price does not reach the buy threshold, no trades are initiated. Similarly, if the stock price does not reach the sell threshold, the stock is held until the price rises to the sell threshold or drops to the buy threshold.

The new algorithm provides several benefits over the previous model. Firstly, it minimizes transaction costs, which in turn reduces the risks associated with frequent trading. Secondly, it minimizes the impact of market volatility by providing a buffer against sudden market movements. Finally, it provides a more stable and predictable investment strategy that is based on historical data and market trends.In conclusion, the new decision algorithm with a threshold provides a more effective and reliable investment strategy that minimizes risks and maximizes profits. Algorithm displayed in Algorithm\ref{al-Xiugai}.

The algorithm\ref{al-Xiugai} which in \ref{al-app} above outlines the algorithmic logic of a stock trading strategy that utilizes machine learning to predict trading signals and execute buy/sell orders based on those signals. The first step is to extract the date information from the historical stock data $df$. Next, the machine learning model is used to predict trading signals, resulting in a trading signal sequence $grid$. The historical stock data is then iterated over, and for each time point, the corresponding buy/sell strategy is executed based on the signals in $grid$.

If the current signal is a buy signal and there is no existing position, a buy order is executed at the current market price, and the buy price and buy time are recorded. The position size $buy\_cishu$ is set to 1. If there is an existing position, the current price is compared to the buy price to determine whether the profit threshold $buy$ has been reached. If it has, more shares are bought, and the position size and buy time are updated. Otherwise, the existing position is held.

If the current signal is a sell signal and there is an existing position, the current price is compared to the buy price to determine whether the selling condition has been met. If it has, the profit is calculated, and the sell order and profit result are recorded in the appropriate list. The buy price and position size are reset to 0. Otherwise, the existing position is held.

If the current signal is a buy signal and there is an existing position, the current price is compared to the buy price to determine whether the position should be increased. If it should be, more shares are bought, and the position size and buy price are updated. Otherwise, the existing position is held.

The above steps are repeated until the last time point in the historical data is reached. The output is a list of recorded profit results. While this is a simple example, actual trading strategies must consider additional factors such as trading costs and liquidity risk, and require careful design and optimization.

\begin{table}[H]
	\centering
	\resizebox{\textwidth}{!}{
		\begin{tabular}{c|cccccccccc|c}
			\toprule
			\rowcolor[gray]{0.8} & ST Yi Paper & Fengdong Co., Ltd. & Transfer of intellectual education & An Naier & Magnificent Blast & Zhaohe Industrial & Fangda Chemical & Xingmei United & Zhejiang Guangxia & Red Bean Co., Ltd. & Avg. \\
			\toprule
			total time(s) & 6.662992716&6.414739609&3.09473896&5.047596931&6.504285097&3.08650589&6.606285334&6.849395514&6.55560565&7.205817938&5.803\\
			total revenue(\%) &0.074702931&0.116811043&-0.266774914&-0.024660367&0.127029218&-0.009518885&0.077444503&-0.10783974&0.225763105&0.154298468&0.0367\\
			train\_accuracy &0.871794872&0.87804878&0.849315068&0.863013699&0.871165644&0.890410959&0.867142857&0.980392157&0.911111111&0.824146982&0.881\\
			test\_accuracy &0.846153846&0.714285714&0.918918919&0.864864865&0.780487805&0.891891892&0.835227273&0.769230769&0.882352941&0.885416667&0.839\\
			\bottomrule
	\end{tabular}}
	\caption{The logic regression with threshold fitting effect of the rise and fall ranges of various stock models}
	\label{tab:log-result-2}
\end{table}

\begin{table}[H]
	\centering
	\resizebox{\textwidth}{!}{
		\begin{tabular}{c|cccccccccc|c}
			\toprule
			\rowcolor[gray]{0.8} & ST Yi Paper & Fengdong Co., Ltd. & Transfer of intellectual education & An Naier & Magnificent Blast & Zhaohe Industrial & Fangda Chemical & Xingmei United & Zhejiang Guangxia & Red Bean Co., Ltd. & Avg. \\
			\toprule
			total time(s) & 12.286 & 8.353 & 2.885 & 6.511 & 8.521 & 2.978 & 8.067 & 7.893 & 8.062 & 7.908 & 7.747 \\[0.5ex]
			total revenue(\%) & 0.604 & 0.528 & 0.264 & 0.145 & 0.416 & 0.233 & 0.835 & 0.570 & 0.570 & 0.143 & 0.406 \\[0.5ex]
			train\_accuracy & 0.643 & 0.678 & 0.609 & 0.703 & 0.642 & 0.578 & 0.695 & 0.645 & 0.594 & 0.656 & 0.632 \\[0.5ex]
			test\_accuracy & 0.588 & 0.707 & 0.564 & 0.527 & 0.617 & 0.536 & 0.7 & 0.591 & 0.595 & 0.582 & 0.597 \\
			\bottomrule
	\end{tabular}}
	\caption{The SVM with threshold fitting effect of the rise and fall ranges of various stock models}
	\label{tab:SVM-result-2}
\end{table}

\begin{table}[H]
	\centering
	\resizebox{\textwidth}{!}{
		\begin{tabular}{c|cccccccccc|c}
			\toprule
			\rowcolor[gray]{0.8} & ST Yi Paper & Fengdong Co., Ltd. & Transfer of intellectual education & An Naier & Magnificent Blast & Zhaohe Industrial & Fangda Chemical & Xingmei United & Zhejiang Guangxia & Red Bean Co., Ltd. & Avg. \\
			\toprule
			total time(s) & 74.42994905 & 73.87512374 & 56.16749215 & 69.80366015 & 77.05178332 & 60.12428188 & 73.07069969 & 72.95501733 & 71.02180123 & 73.11442947 & 70.311 \\[0.5ex]
			total revenue(\%) & 1.19837024 & 0.845676093 & 1.016574871 & 0.86925183 & 0.597742433 & 1.021417206 & 1.081817462 & 1.382396366 & 1.049963555 & 0.978267758 & 0.994 \\[0.5ex]
			train\_accuracy & 0.856164384 & 0.856164384 & 0.849315068 & 0.818493151 & 0.876712329 & 0.815068493 & 0.866438356 & 0.839041096 & 0.815068493 & 0.815068493 & 0.837 \\[0.5ex]
			test\_accuracy & 0.878378378 & 0.837837838 & 0.837837838 & 0.716216216 & 0.837837838 & 0.797297297 & 0.837837838 & 0.864864865 & 0.864864865 & 0.864864865 & 0.834 \\
			\bottomrule
	\end{tabular}}
	\caption{The FCNN with threshold fitting effect of the rise and fall ranges of various stock models}
	\label{tab:FCNN-result-2}
\end{table}

The table\cref{tab:log-result-2,tab:SVM-result-2,tab:FCNN-result-2} in the table represents the performance of the decision model with transaction quantity restrictions. Compared to the previous model, we can see that the modified model with transaction limits has made significant improvements in terms of total revenue. The percentage of total revenue increased from an average of 0.111\% in the original model to 0.191\% in the modified model. This indicates that compared to the original model, models with trading limits can bring more profits to investors.
However, this improvement comes at the cost of increasing the total time required for computation. The average total time of the modified model is 1.75 seconds, which is nearly 50\% longer than the average 1.24 seconds of the original model. This increase in computational time may be due to the increased complexity of transaction constrained algorithms. But in China's daily trading market, one day is enough time to make decisions. In terms of model accuracy, the accuracy levels of the two models in training and testing are similar, and the testing accuracy is equivalent.
Overall, although modified models with trading restrictions can bring more profits to investors, they also require more computational time. It is important to consider the trade-off between profitability and the risks associated with frequent trading when deciding which model to use. If we want to achieve a profit of 140\% of the original model's CNN, we will have to bear the risk of significant transaction fees.

\begin{table}[H]
	\centering	\resizebox{\textwidth}{!}{
		\begin{tabular}{c|cccccccccc}
			\toprule
			\rowcolor[gray]{0.8}& ST Yi Paper & Fengdong Co., Ltd. & \thead{Transfer of \\intellectual education} & An Naier & Magnificent Blast & Zhaohe Industrial & Fangda Chemical & Xingmei United & Zhejiang Guangxia & Red Bean Co., Ltd. \\
			\toprule
			\thead{Number of buy transactions\\ in log model} & 1 & 50 & 23 & 98 & 74 & 78 & 1 & 15 & 1 & 40\\
			\thead{Number of sell transactions\\ in log model} & 0 & 31 & 17 & 65 & 61 & 73 & 0 & 10 & 0 & 37\\
			\rowcolor[gray]{0.8}\thead{Total number of transactions\\ in log model}  & 1 & 81 & 40 & 163 & 135 & 151 & 1 & 25 & 1 & 77 \\
			\thead{Number of buy transactions \\in SVM model}& 16 & 51 & 67 & 41 & 17 & 50 & 24 & 34 & 37 & 33  \\
			\thead{Number of sell transactions\\ in SVM model }& 9 & 35 & 22 & 30 & 11 & 29 & 16 & 22 & 25 & 22  \\
			\rowcolor[gray]{0.8}\thead{Total number of transactions\\ in SVM model} & 25 & 86 & 89 & 71 & 28 & 79 & 40 & 56 & 62 & 55 \\
			\thead{Number of buy transactions\\ in FCNN model} & 107 & 97 & 91 & 41 & 87 & 87 & 90 & 99 & 96 & 99\\
			\thead{Number of sell transactions\\ in FCNN model} & 61 & 60 & 56 & 31 & 46 & 67 & 68 & 67 & 59 & 55\\
			\rowcolor[gray]{0.8}\thead{Total number of transactions\\ in FCNN model}& 168 & 157 & 147 & 72 & 133 & 154 & 158 & 166 & 155 & 154 \\
			\bottomrule
	\end{tabular}}
	\caption{Table of Decision Times after Setting Thresholds}
	\label{tab:Table of Decision Times after Setting Thresholds}
\end{table}

From the table\ref{tab:Table of Decision Times after Setting Thresholds}, it can be seen that the updated model includes the ability to limit the number of transactions, thereby reducing the risks associated with frequent transactions. Let's analyze the results of the model from the perspective of risk and return. Firstly, let's take a look at the returns generated by the model. The log model has the highest number of transactions, which may bring higher returns. However, an excessive number of transactions can also increase transaction costs and potentially lead to a decrease in net return. In contrast, the SVM model has a relatively low number of transactions, which can reduce transaction costs but may also reduce the potential for returns. The FCNN model has the highest number of buying transactions, which indicates that market sentiment is bullish. A large number of buying transactions indicate that the model is confident in the market direction and expects stock prices to rise in the future. However, the large number of transactions also increases the risk associated with frequent transactions. Secondly, let's analyze the risks involved in the model. The log model has the highest number of transactions, which increases the risk associated with frequent transactions. The SVM model has a relatively low number of transactions, which can reduce transaction costs but may also have the potential to reduce returns.

\begin{figure}[H]
	\centering
	\includegraphics[width=0.45\linewidth]{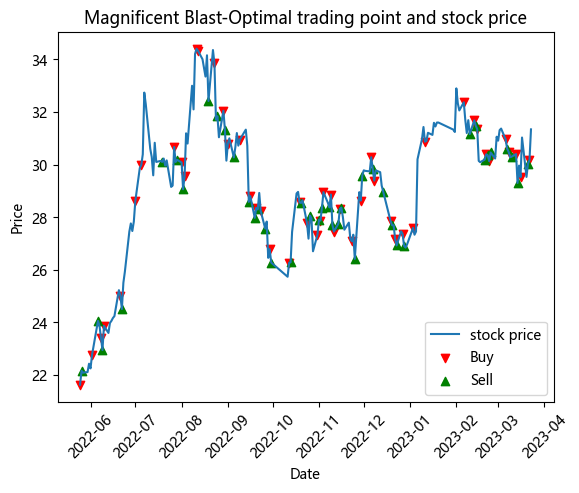}
	\includegraphics[width=0.45\linewidth]{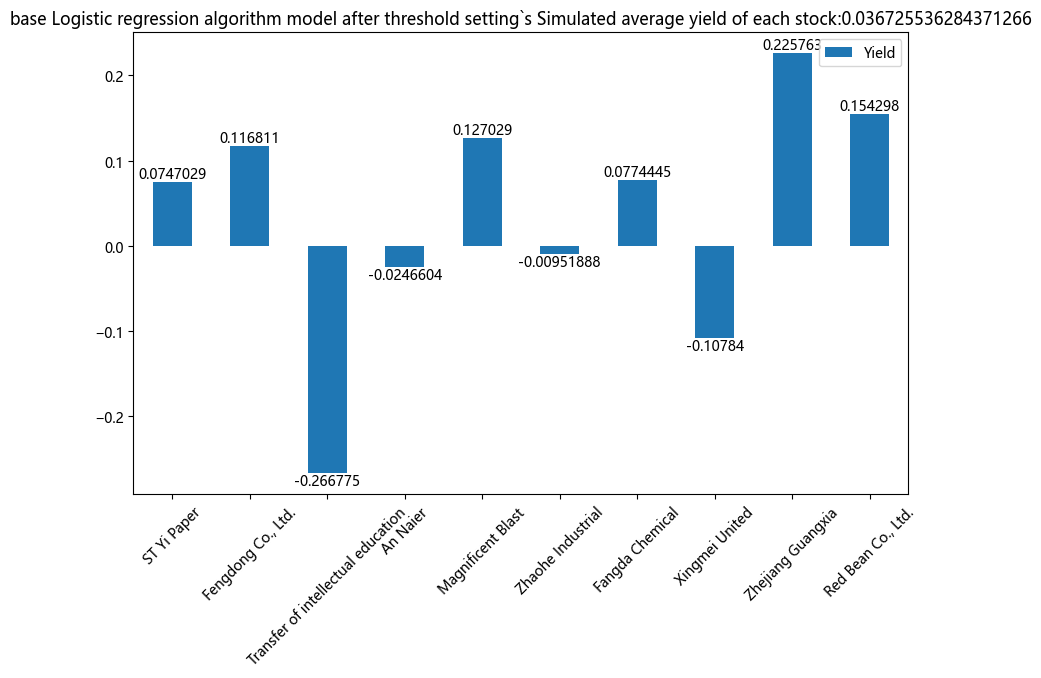}
	\caption{Partial results display using a logic regression model with added thresholds as an example}
	\label{result-after}
\end{figure}

\begin{figure}[H]
	\centering
	\includegraphics[width=0.31\linewidth]{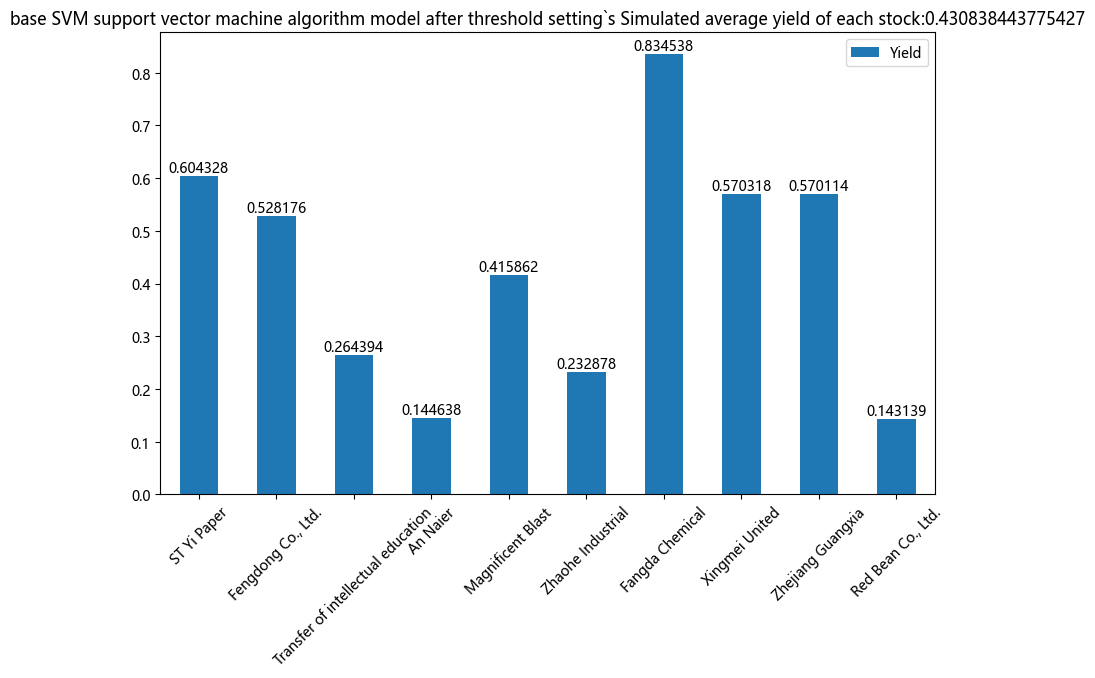}
	\includegraphics[width=0.31\linewidth]{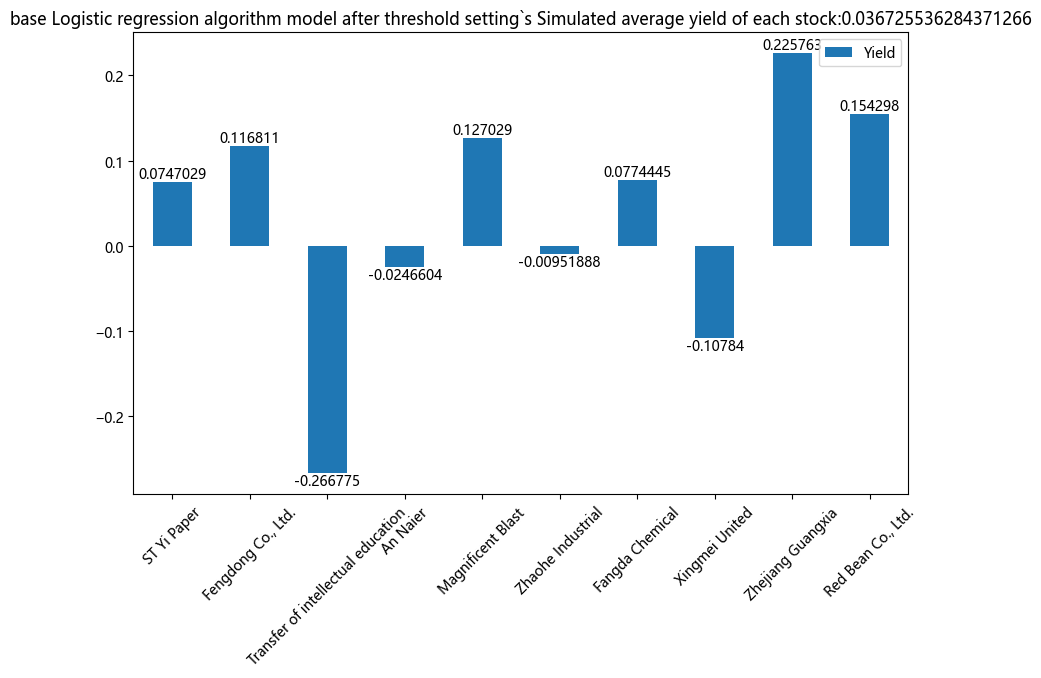}
	\includegraphics[width=0.31\linewidth]{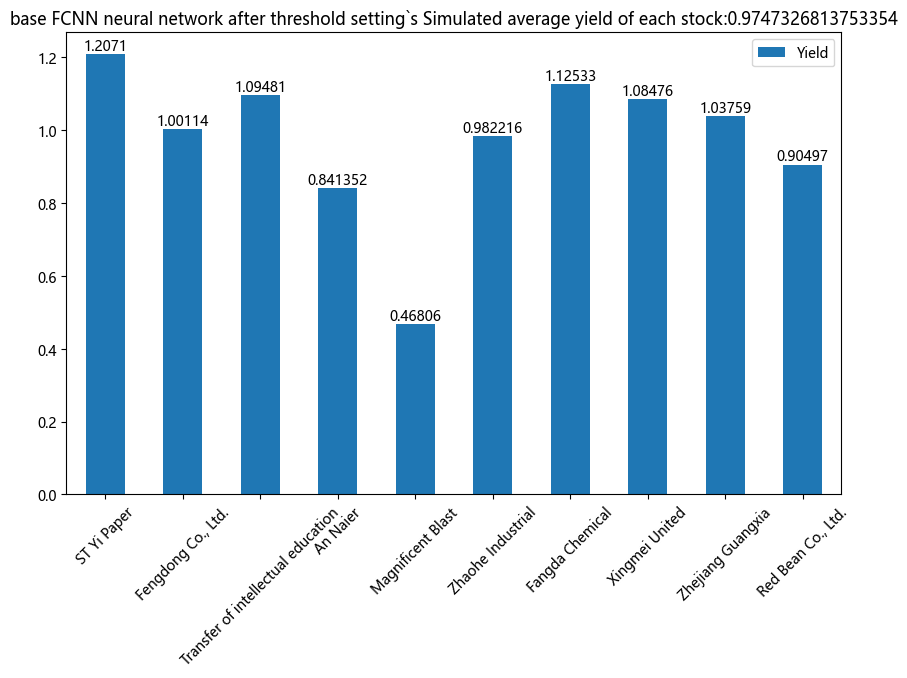}
	
	\caption{Ten stock average return charts without set thresholds based on different models}\label{after-aver}
\end{figure}

\begin{figure}[H]
	\centering
	\includegraphics[width=0.18\linewidth]{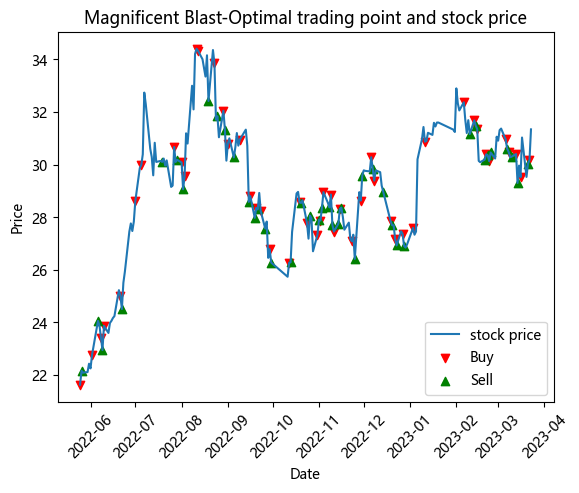}
	\includegraphics[width=0.18\linewidth]{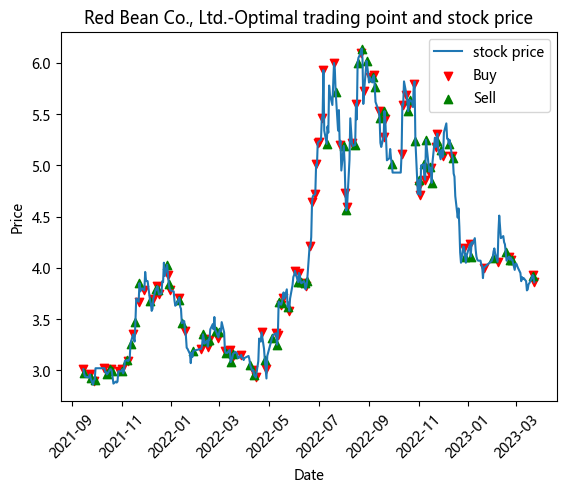}
	\includegraphics[width=0.18\linewidth]{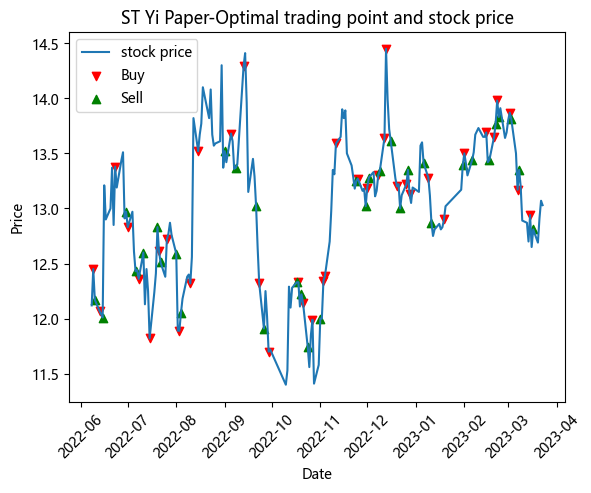}
	\includegraphics[width=0.18\linewidth]{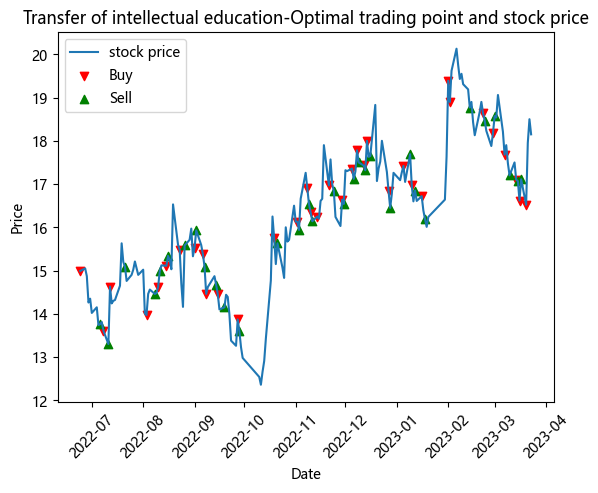}
	\includegraphics[width=0.18\linewidth]{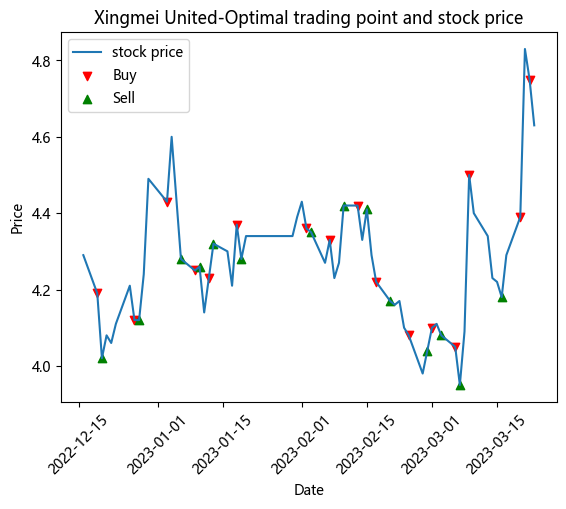}
	\\
	\includegraphics[width=0.18\linewidth]{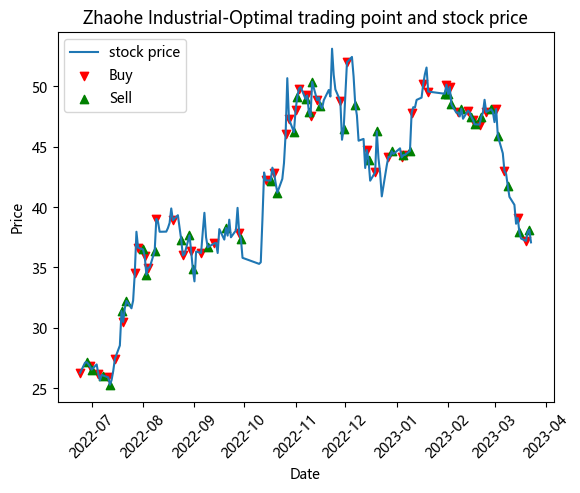}
	\includegraphics[width=0.18\linewidth]{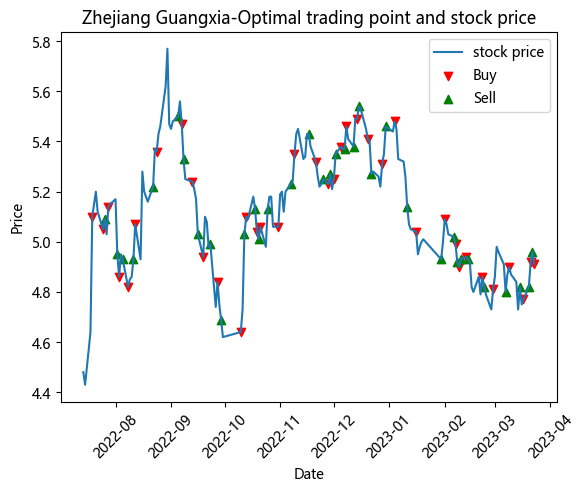}
	\includegraphics[width=0.18\linewidth]{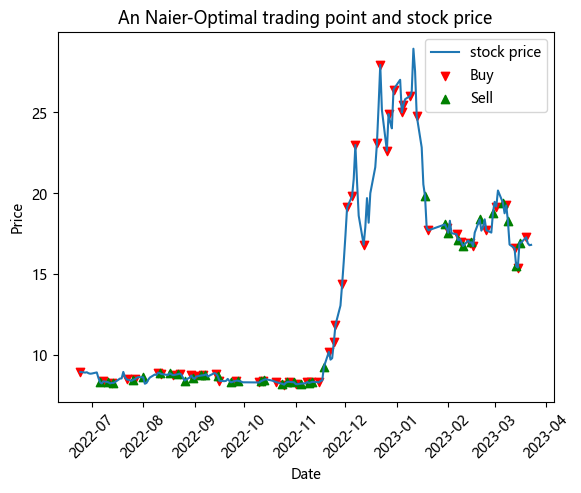}
	\includegraphics[width=0.18\linewidth]{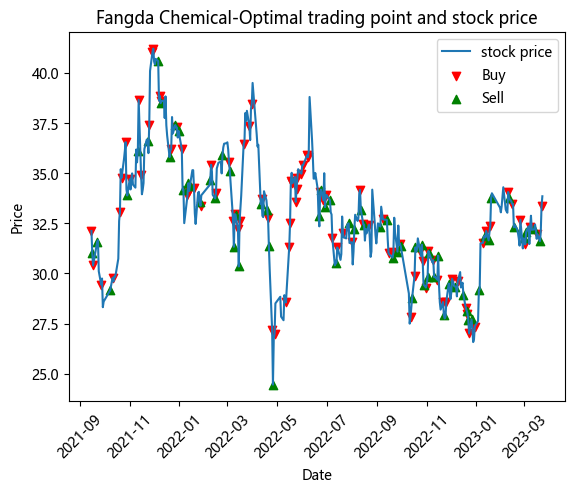}
	\includegraphics[width=0.18\linewidth]{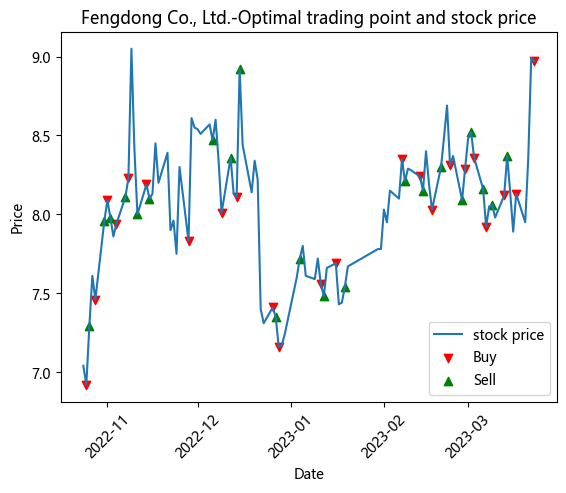}
	\caption{after setting threshold logic regression decision results}
	\label{B1}
\end{figure}

\begin{figure}[H]
	\centering
	\includegraphics[width=0.18\linewidth]{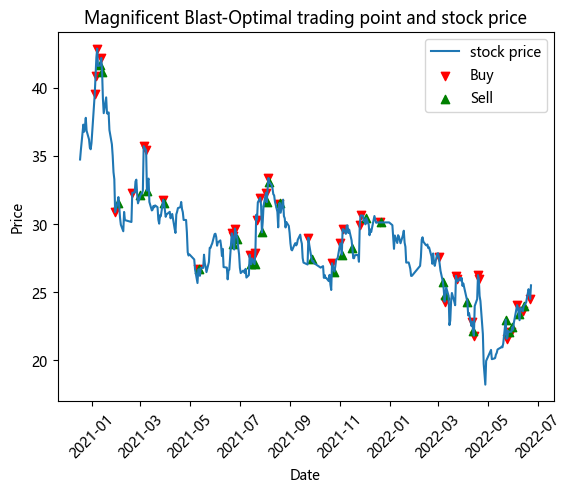}
	\includegraphics[width=0.18\linewidth]{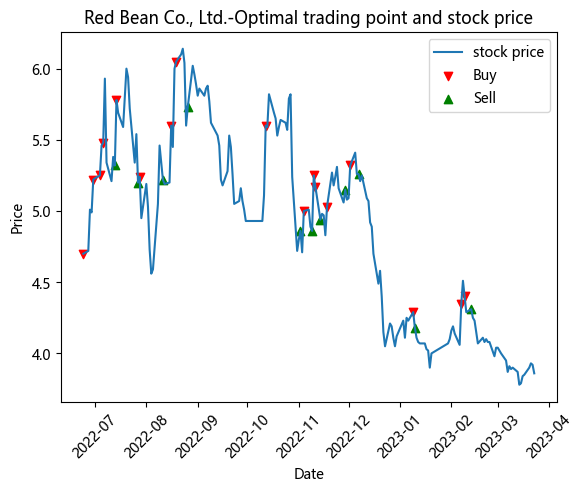}
	\includegraphics[width=0.18\linewidth]{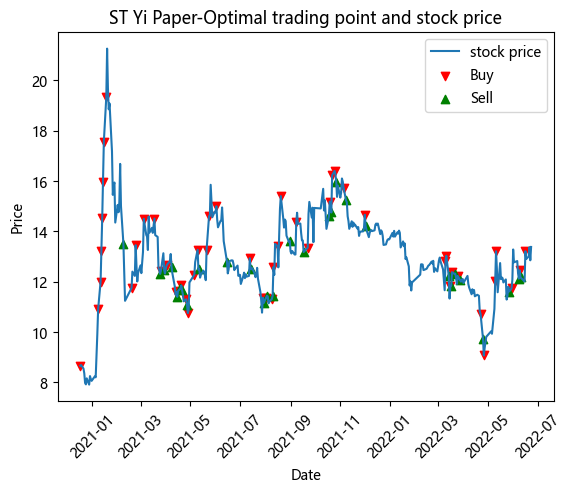}
	\includegraphics[width=0.18\linewidth]{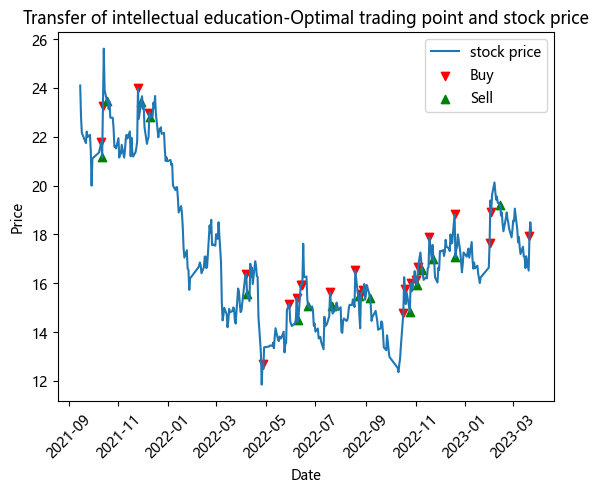}
	\includegraphics[width=0.18\linewidth]{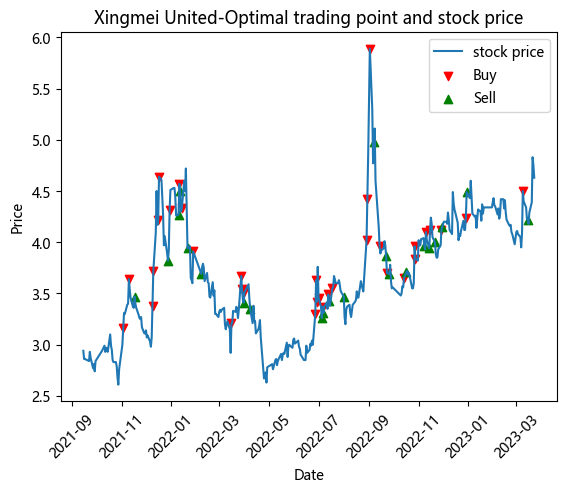}
	\\
	\includegraphics[width=0.18\linewidth]{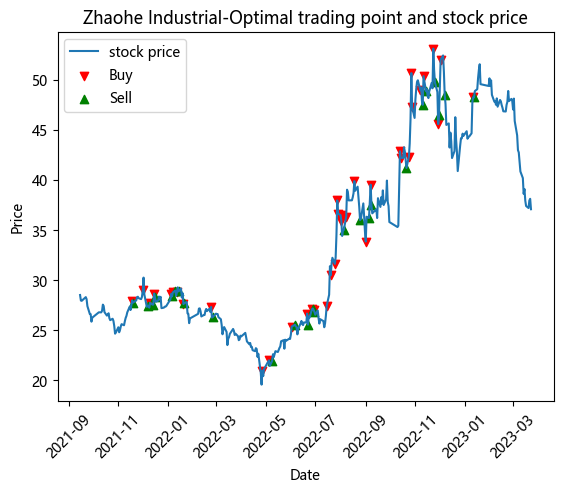}
	\includegraphics[width=0.18\linewidth]{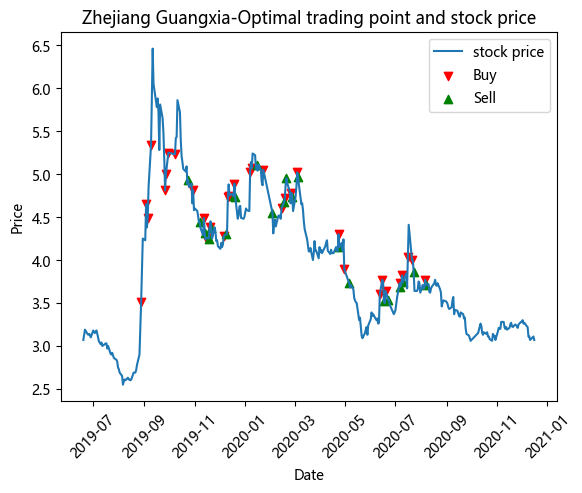}
	\includegraphics[width=0.18\linewidth]{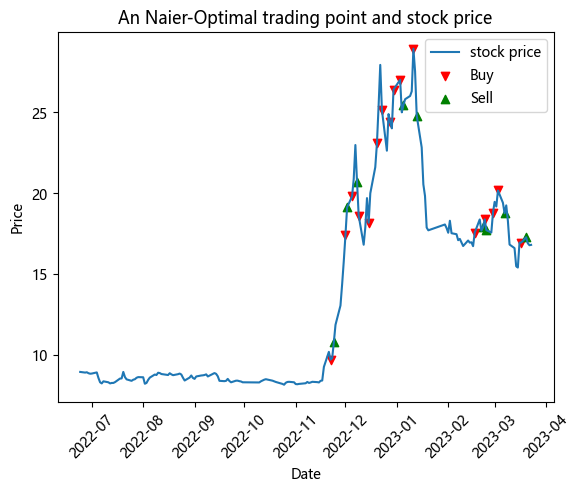}
	\includegraphics[width=0.18\linewidth]{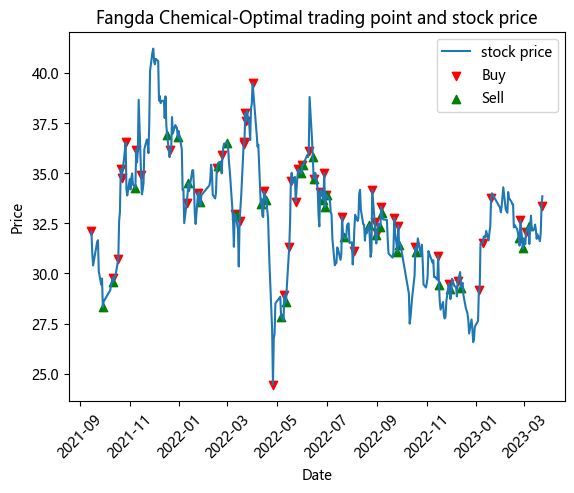}
	\includegraphics[width=0.18\linewidth]{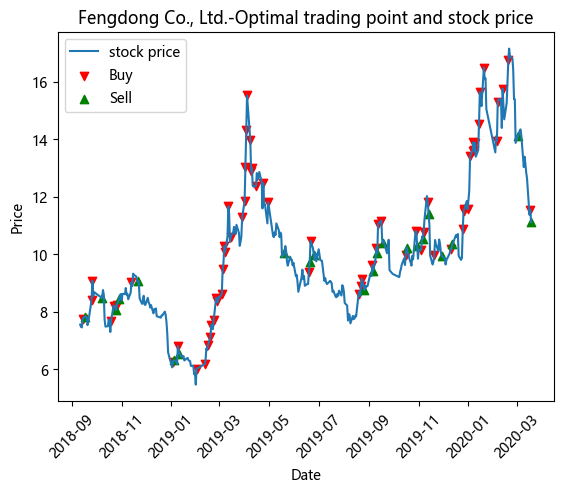}
	\caption{after setting threshold SVM decision results}
	\label{B2}
\end{figure}

\begin{figure}[H]
	\centering
	\includegraphics[width=0.18\linewidth]{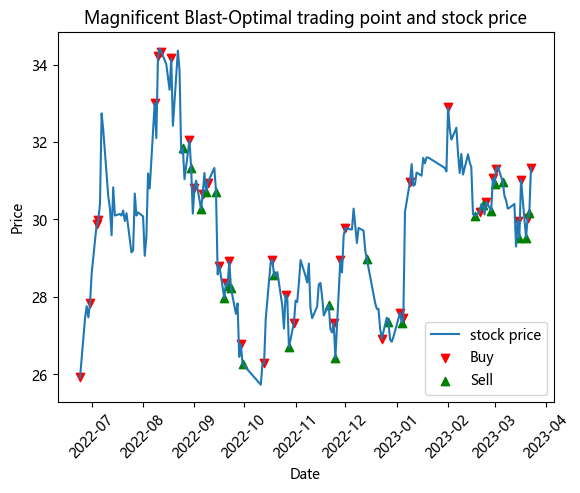}
	\includegraphics[width=0.18\linewidth]{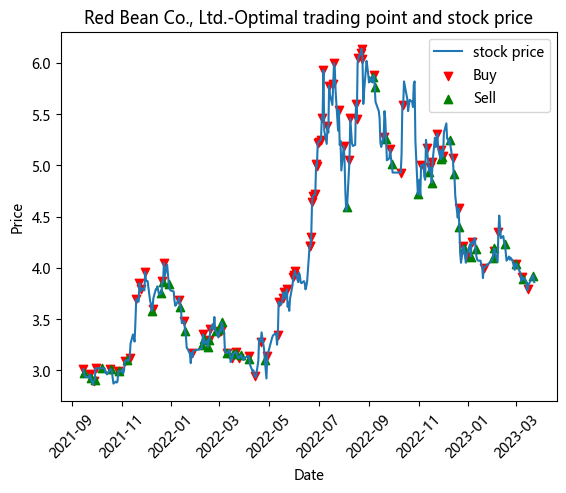}
	\includegraphics[width=0.18\linewidth]{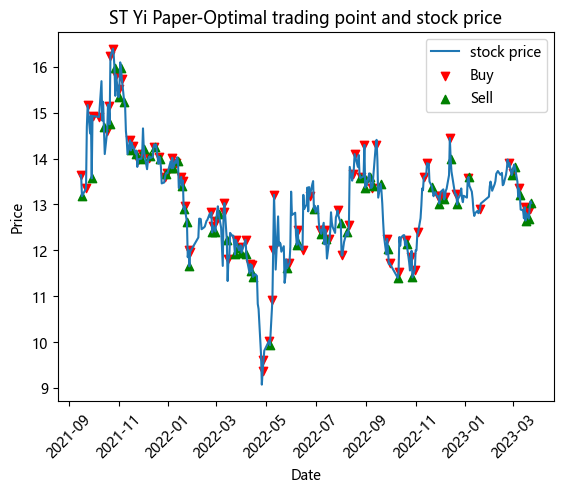}
	\includegraphics[width=0.18\linewidth]{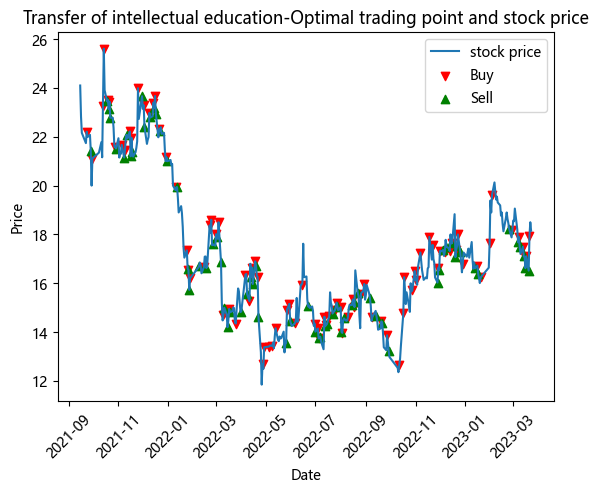}
	\includegraphics[width=0.18\linewidth]{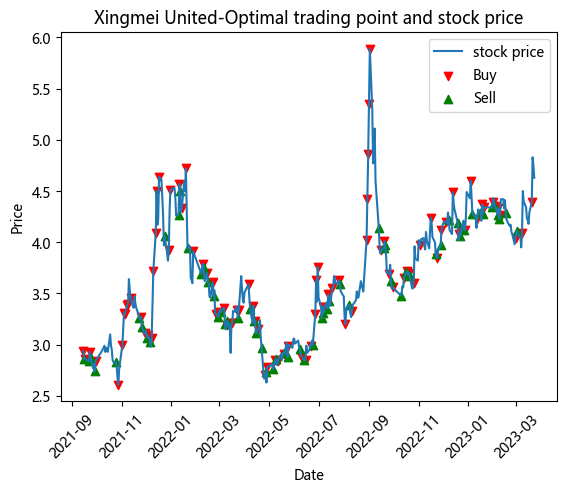}
	\\
	\includegraphics[width=0.18\linewidth]{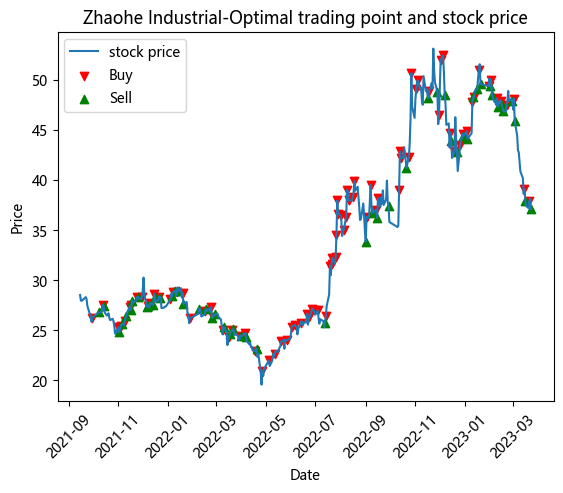}
	\includegraphics[width=0.18\linewidth]{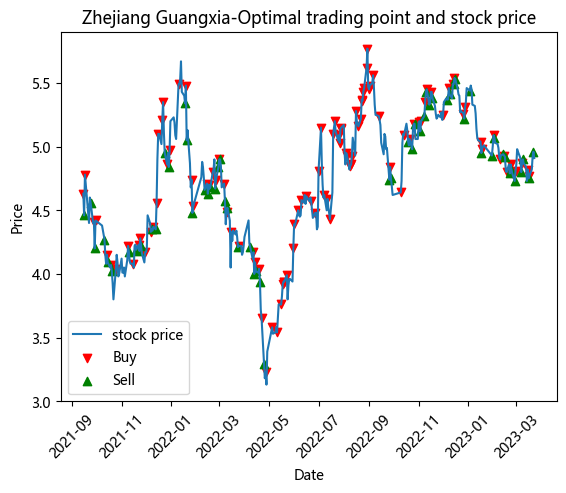}
	\includegraphics[width=0.18\linewidth]{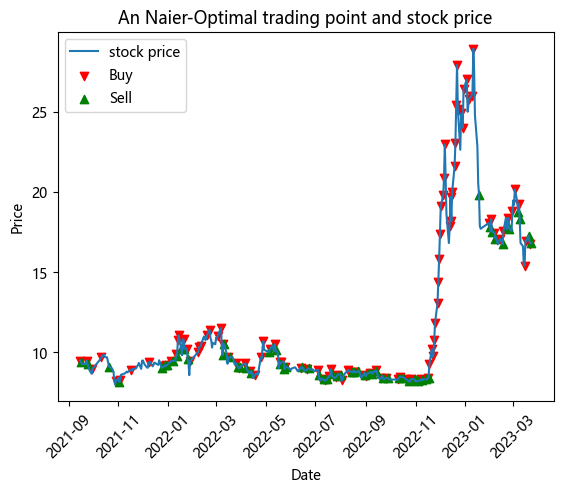}
	\includegraphics[width=0.18\linewidth]{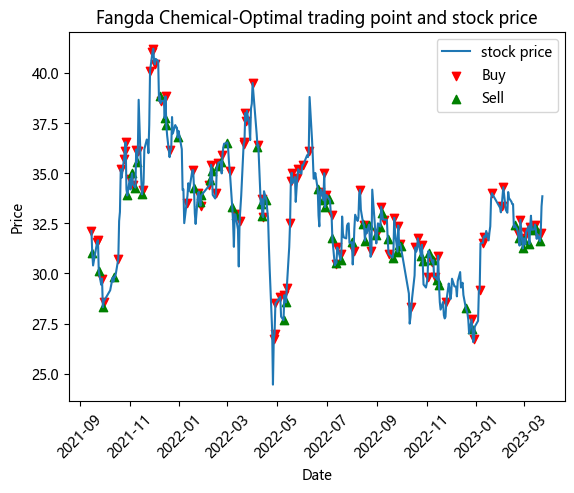}
	\includegraphics[width=0.18\linewidth]{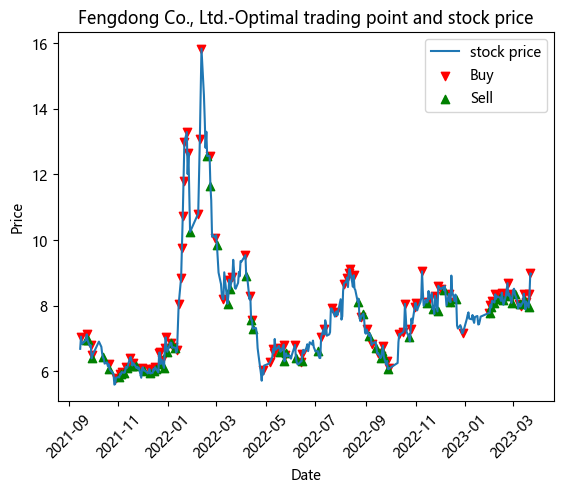}
	\caption{after setting threshold FCNN decision result}
	\label{B3}
\end{figure}

From the figure\ref{result-after}, it can be seen that the limited transaction algorithm implemented in the new model reduces the number of transactions, thereby reducing the risks associated with frequent transactions. Compared with the original model, the updated model has lower transaction costs and risks associated with frequent transactions. However, a decrease in transaction volume may also limit the potential for returns. It is important to strike a balance between the potential returns and risks of frequent transactions, and the updated model seems to be achieving this. Investors should consider their risk tolerance and investment goals before choosing the model to follow.In order to save space in the main text, this is only presented as the visualization results of logic regression. (The visualization results of the three optimized models for 10 companies are shown in \cref{after-aver,B1,B2,B3}.)

\section{Conclusion}

In recent years, machine learning algorithms have been widely applied in the financial field to develop decision models for financial market transactions. This has led to the development of various models, such as logarithmic model, support vector machine (SVM) and feedforward neural network (FCNN). In this study, we analyzed the performance of these three models in predicting buying and selling transactions of ten different stocks.

Analysis shows that the FCNN model outperforms other models in terms of risk and return. Compared with other models, the FCNN model has a higher return on investment and lower risk of loss. The FCNN model predicts more purchase transactions than other models, resulting in higher returns. By optimizing the model, the number of transactions will be reduced, which leads to a reduction in the risk of loss caused by transaction costs.
From the perspective of model optimization. The increase in transaction count restrictions resulted in a significant decrease in the number of transactions in all three models. However, in terms of risk and return, the FCNN model still outperforms other models.

The results of this study indicate that the FCNN model may be a valuable tool for investors who wish to make trading decisions in the financial market. Compared with other models, this model provides a higher investment return rate and lower loss risk. However, it should be noted that these models are not foolproof, as there is always a risk of loss in financial markets.

In addition, the study emphasizes the importance of considering transaction costs when making trading decisions. The restriction on the number of transactions has led to a significant reduction in transaction costs, which may have a significant impact on the overall investment return. Therefore, investors should carefully consider transaction costs when making trading decisions. In summary, this study demonstrates the effectiveness of machine learning algorithms in developing financial market transaction decision models. The FCNN model outperforms other models in terms of risk and return, and the limitation on the number of transactions leads to a reduction in transaction costs. However, investors should always be aware of the risks involved in financial markets and carefully consider transaction costs when making trading decisions.
\newpage
\bibliographystyle{elsarticle-num} 

\bibliography{reference}
\appendix
\newpage
\section{Decision Algorithm}\label{al-app0}
\begin{algorithm}[H]
	\SetAlgoLined
	\KwIn{y\_pred: predicted signals, df\_selected: selected stock data}
	\KwOut{profits: profit for each trading, strategies: trading strategies}
	Initialize buy\_price to None\;
	Initialize profits and strategies lists to empty\;
	\For{each row in df\_selected}{
		\If{y\_pred[i-len(dddate)+Longs] == 1}{
			Set buy\_price to today's open price\;
			Set buy\_index to current index\;
			Set buy\_cishu to 1\;
			Append 0 to profits\;
			Append "buy" to strategies\;
		}
		\Else{
			\If{buy\_price is not None}{
				Calculate profit as (today's open price * buy\_cishu - buy\_price) / (buy\_price)\;
				Append profit to profits\;
				Append "sell" to strategies\;
				Set buy\_price to None\;
				Set buy\_cishu to 0\;
			}
			\Else{
				Append 0 to profits\;
				Append "hold" to strategies\;
			}
		}
	}
	\caption{Buy and Sell Trading Strategy}
\end{algorithm}
\newpage
\section{Decision Algorithm with Threshold}\label{al-app}
\begin{algorithm}[H]
	\footnotesize
	\SetAlgoLined
	\SetAlgoLongEnd 
	\KwIn{Historical stock data $df$, $Longs$, $Shorts$, $buy$, $sell$}
	\KwOut{Profits from trading strategy}
	Extract dates from $df$ and predict trading signals using ML to get $grid$\;
	Initialize $buyPrice$, $buyIndex$, $profits$, $strategies$ as None, None, [], []\;
	
	\For{$i\gets0$ \KwTo $len(df)-1$}{
		$row \gets df.iloc[i]$\;
		\If{$grid[i-len(dates)+Longs] = 1$}{
			\If{$buyPrice =$ None}{
				$buyPrice \gets row['Open']$, $buyIndex \gets i$, $profits$.append(0), $strategies$.append('Buy')\;
			}
			\ElseIf{$(row['Open']*buy\_cishu - buyPrice) / buyPrice > buy$}{
				$buyPrice \gets buyPrice+row['Open']$, $buyIndex \gets i$, $buy\_cishu\gets buy\_cishu+1$, $profits$.append(0), $strategies$.append('Buy')\;
			}
			\Else{$profits$.append(0), $strategies$.append('Hold')}
		}
		\Else{
			\If{$buyPrice \neq$ None}{
				\If{$(row['Open']*buy\_cishu - buyPrice) / buyPrice < sell$}{
					$profits$.append((row['Open']*buy\_cishu - buyPrice) / buyPrice), $strategies$.append('Sell'), $buyPrice\gets$ None, $buy\_cishu\gets 0$\;
				}
				\Else{$profits$.append(0), $strategies$.append('Hold')}
			}
			\Else{$profits$.append(0), $strategies$.append('Hold')}
		}
	}
	
	\Return{$profits$}
	\caption{Decision Algorithm with Threshold}
	\label{al-Xiugai}
\end{algorithm}

\end{document}